\documentclass[12pt,preprint]{aastex}
\received{}
\accepted{}
\slugcomment{}
\shorttitle{FUV Ups and Downs of $\alpha$~Cen}
\shortauthors{Ayres}

\begin{document}

\title{The Far-Ultraviolet Ups and Downs of Alpha Centauri}

\author{Thomas R.\ Ayres}

\affil{Center for Astrophysics and Space Astronomy,\\
389~UCB, University of Colorado,
Boulder, CO 80309;\\ Thomas.Ayres@Colorado.edu}

\begin{abstract}

Four years (2010--2014) of semiannual pointings by {\em Hubble}\/ Space Telescope Imaging Spectrograph (STIS) on  nearby Alpha Centauri have yielded a detailed time history of far-ultraviolet (FUV: 1150--1700~\AA) emissions of the solar-like primary (A: G2~V) and the cooler, but more active, secondary (B: K1~V).  This period saw A climbing out of a prolonged coronal X-ray minimum, as documented contemporaneously by {\em Chandra,}\/ while B was rising to, then falling from, a peak of its long-term ($\sim$~8 yr) starspot cycle.  The FUV fluxes of the primary were steady over most of the STIS period, although the [\ion{Fe}{12}] $\lambda$1242 coronal forbidden line ($T\sim 1.5$~MK) partly mirrored the slowly rising X-ray fluxes.  The FUV emissions of the secondary more closely tracked the rise and fall of its coronal luminosities, especially the ``hot lines'' like \ion{Si}{4}, \ion{C}{4}, and \ion{N}{5} ($T\sim 0.8$--$2\times10^{5}$~K), and coronal [\ion{Fe}{12}] itself.  The hot lines of both stars were systematically redshifted, relative to narrow chromospheric emissions, by several km s$^{-1}$, showing little change in amplitude over the 4-year period; especially for $\alpha$~Cen B, despite the significant evolution of its coronal activity.  Further, the hot lines of both stars, individually and epoch-averaged, displayed non-Gaussian shapes, which most trivially could be decomposed into two components, one narrow (${\rm FWHM}\sim 25$--45 km s$^{-1}$), the other broad (60--80~km s$^{-1}$).  The bimodal Gaussian strategy had been applied previously to the $\alpha$~Cen stars, but this was the first opportunity to evaluate any time dependence.  In fact, not much variation of the component properties was seen, even over the major cycle changes of B.  Curiously, the line fluxes were about equally divided between the narrow and broad components for both stars.  The fact that there is minimal activity-dependence of the narrow/broad flux partition, as well as densities derived from \ion{O}{4}] line ratios, either during the cycle evolution of B, or between A and B, suggests that there is a dominant ``quantum'' of FUV surface activity that is relatively unchanged during the cycle, aside from the fractional area covered.
\end{abstract}


\section{INTRODUCTION}

The nearby $\alpha$~Centauri triple system, especially the central AB binary (A: G2~V; B: K1~V), is a keystone for understanding the properties of magnetically inspired ``activity'' on sun-like stars of solar age.  The two companions are bright in all the high-energy bands, from X-rays to the ultraviolet, that are important diagnostics of the pronounced outer-atmosphere response to the activity.  The system is one of very few that has numerous epochs of coronal ($T\sim 1$~MK) X-ray observations, dating back more than thirty years, including significant attention over the past decade from {\em XMM-Newton}\/ (Robrade et al.\ 2012) and {\em Chandra}\/ (Ayres 2014).  The present study considers a unique set of far-ultraviolet (1150--1700~\AA) pointings on $\alpha$~Cen AB by {\em Hubble}\/ Space Telescope Imaging Spectrograph (STIS) over the past four years, on a semi-annual basis, in support of a longer-term {\em Chandra}\/ X-ray program.  

Beech (2012) has provided a recent comprehensive review of the properties of the $\alpha$~Centauri system, and a remarkable history of astronomical observations extending back to the nineteenth century.  Especially germane to the present effort, Pagano et al.\ (2004) carried out a detailed high-resolution STIS program on specifically $\alpha$~Cen A, including extensive line identifications and profile fitting.  More recently, Linsky et al.\ (2012) have described STIS and {\em Hubble}\/ Cosmic Origins Spectrograph (COS) FUV measurements of fast-rotating, super-active sun-like stars, incorporating the Pagano et al.\ spectra of $\alpha$~Cen A as a low-activity comparison.  These two studies provide complementary contexts for the multi-epoch STIS program here.

\section{OBSERVATIONS}

The FUV spectra of AB were collected as part of a joint {\em Chandra/HST}\/ project, beginning 2010 after STIS was repaired in Servicing Mission 4 (SM4).  The original, {\em Chandra-}only part of the program was instigated in late-2005, in response to a report from {\em XMM-Newton}\/ of an unprecedented drop in $\alpha$~Cen A's 0.2--2~keV soft X-ray luminosity (Robrade et al.\ 2005), so-called by the authors the ``darkening of the solar twin'' or the ``fainting of $\alpha$~Cen A.''  The {\em Chandra}\/ observations from 2005 onward confirmed that A was in a coronal low state with respect to its behavior during the {\em ROSAT}\/ period in the 1990's, but the depth of the decline was much shallower than suggested by {\em XMM-Newton}\/ (see, e.g., Ayres 2014, and references to previous work therein).

\subsection{{\em Chandra}\/ High-Resolution Camera}

The {\em Chandra}\/ High-Resolution Camera (HRC) imaging of $\alpha$~Cen AB up to mid-2013 has been described by Ayres (2014), where details concerning the instrument configurations and data analysis can be found.  Since that study, two additional {\em Chandra}\/ pointings have been carried out, in late-2013 and mid-2014.  These are summarized in Table~1.  The new $L_{\rm X}/L_{\rm bol}$ ratios, together with the previous time series, are illustrated in Figure~1.  

The solar values in Fig.~1 are 81~day averages (three rotations), over Cycles 23 and (beginning of) 24; error flags are 1\,$\sigma$ standard deviations of the daily measurements in each 81~day bin, a rough measure of temporal variability.  A three-cycle average is shaded gray, projected into the future as dashed curves, highlighting the unusual extended minimum of Solar Cycle 23.  Larger solid dots represent X-ray fluxes of $\alpha$~Cen AB from the {\em Chandra}\/ HRC pointings (Y2000 and later) and four earlier epochs from the {\em ROSAT}\/ High-Resolution Imager.  Asterisks are published {\em XMM-Newton}\/ X-ray luminosities of B (A cannot easily be resolved from brighter B since about 2006, owing to the shrinking orbital separation in recent years) scaled slightly to match the apparent {\em Chandra}\/ variation.  Dot-dashed curves are schematic log-sinusoidal fits to the AB cycles ($P\sim 19$~yr, 8~yr, respectively).  Triangles mark epochs when STIS FUV echelle spectra were taken.

The latest $L_{\rm X}/L_{\rm bol}$ values confirm the recent generally downward trend of the secondary star, but upward trajectory of the primary, which had been mired in a long-term low state during 2005--2010, as mentioned earlier.

\subsection{Space Telescope Imaging Spectrograph}

The characteristics and operational capabilities of STIS have been described in a number of previous publications, especially Woodgate et al.\ (1998) and Kimble et al.\ (1998).

The STIS pointings in 2010--2014 were during a period when the 80 year AB orbit was closing rapidly toward a minimum separation of $\sim 4^{\prime\prime}$ in 2016.  The proximity of these optically very bright stars required a special target acquisition.  First, the brighter of the two, A, was captured with a normal CCD imaging acquisition in visible light through the ND5 filter.  According to the STIS Instrument Handbook\footnote{see: http://www.stsci.edu/hst/stis/documents/handbooks/currentIHB/cover.html}, the CCD acquisition has a 2\,$\sigma$ centering uncertainty of 0.01$^{\prime\prime}$, corresponding to a velocity error of about 1~km s$^{-1}$.  To aid the initial acquisition, the position and proper motion of A were updated via the long-term {\em Chandra}\/ imaging, agreeing best with the historical Luyten (1976) value for the proper motion part.  After the STIS echelle exposures of A were taken, a blind offset was performed to the predicted location of the secondary star.  Here, also, the relative orbit had been refined based on the recent {\em Chandra}\/ positional measurements.  In the early years of the joint program, 2010--2011, the offset was truly blind, to drop B directly into the STIS photometric aperture ($0.2^{\prime\prime}{\times}0.2^{\prime\prime}$).  Because the precision of the relative position was better than $0.1^{\prime\prime}$, the blind offset should have accurately centered the target.  The idea was to capture the secondary star without risking a normal CCD acquisition, which possibly could default to the optically brighter of the two companions (A).  

Unfortunately, the simple blind offset strategy failed in the initial pointing on B, about eight months after the repair of STIS in SM4.  A significant technical delay in the first observation caused the offset embedded in the Phase~II observing script to be out of date with respect to the true relative position, and a table of time-dependent values in the Phase~II specifically for that contingency had been overlooked by the schedulers.  Despite the initial failed pointing, a re-observation of B was carried out successfully less than six months later using the direct offset strategy, but with the epoch-appropriate values.  Subsequently, to avoid any future acquisition issues, a peak-up (search and centering) was conducted following the blind offset, in dispersed visible light with the $0.3{\times}0.05{\rm ND3}$ aperture, whose standard raster pattern sweeps out a roughly $0.3^{\prime\prime}{\times}0.3^{\prime\prime}$ area.

All the STIS $\alpha$~Cen observations were carried out in {\em HST}'s Continuous Viewing Zone (CVZ), which the high-(southern)-declination system falls into a number of times during the year, compatible with the desired six-month cadence of the {\em Chandra}\/ time series.  Two orbits were allocated to each semiannual visit.  For the post-2012 STIS program, which has focused exclusively on E140M-1425 medium-resolution ($R= \lambda/\Delta\lambda\sim 4{\times}10^4$ [$\sim 7$~km s$^{-1}$ FWHM]) FUV spectroscopy, nearly 4.3~ks of exposure could be collected for each star, divided into three equal sub-exposures (to mitigate short-term instrumental drifts due to telescope ``breathing'') in the high-efficiency CVZ.  Table~2 provides a summary of the post-SM4 STIS pointings on AB.

Figure~2 illustrates raw E140M echellegrams of $\alpha$~Cen AB, summed over the nine epochs of the program to emphasize some of the weaker spectral structure.  Shorter wavelengths ($\lambda< 1200$~\AA) are at the bottom, increasing toward the top, and left to right in each echelle order (darker stripes), reaching $\lambda\sim 1700$~\AA\ at the upper edge of the frame.  The gray scale for B saturates at one third the value for A, compensating for the 3 times lower bolometric flux of the smaller, cooler K dwarf.  The intensities in the image strip between $y= 125$--210 pixels were reduced a factor of 10 to allow detail in the bright \ion{H}{1} Ly$\alpha$ line to be seen. 

Prominent emissions are marked in the lower center legend.  Note the duplication of several bright lines (e.g., Ly$\alpha$ red wing, \ion{O}{1} $\lambda$1302) in adjacent orders.  The main difference between the two stars is the rising continuum toward longer wavelengths in A, and presence of conspicuous [\ion{Fe}{12}] $\lambda$1242 coronal forbidden line emission in B (just shortward of bright \ion{N}{5} $\lambda$1242).

\subsubsection{Reduction of STIS Echellegrams}

The STIS spectrograms initially were processed through the On-the-Fly {\sf calstis}\/ pipeline at the Mikulski Archive for Space Telescopes (MAST), yielding the standard ``x1d'' file, a tabulation of spectral properties order-by-order for the particular echelle setting.  An E140M visit to either of the stars was divided into 2--3 sub-exposures, each of which yielded a separate x1d dataset.  The echelle orders in a given sub-exposure then were merged according to protocols developed for the Advanced Spectral Library Project (ASTRAL\footnote{see: http://casa.colorado.edu/$\sim$ayres/ASTRAL/}), to produce a coherent spectral tracing: flux densities, photometric errors, and data quality flags versus wavelength.  One crucial feature of the post-pipeline step was to apply a correction for small-scale wavelength distortions identified in an earlier extensive study of the STIS wavelengths, based on deep exposures of the onboard calibration lamps (Ayres 2008, 2010).  A second key aspect involved an adjustment of the echelle blaze curves to minimize flux discontinuities over adjacent echelle orders.  Unfortunately, this was not feasible for $\alpha$~Cen B owing to insufficient continuum signal to balance the flux densities in the order-overlap zones.  

Next, the sub-exposures of a given observation were registered to the first in velocity, by cross-correlating against the strong, narrow \ion{O}{1} $\lambda$1306 feature; and scaling to the first in flux, by considering ratios between high-S/N regions of each dataset.  The assumption was that the sub-exposures of an observation all should have the same flux density distributions, because the short duration of the visit mitigates against source variability, and any apparent differences must be caused by changes in the effective throughput, say due to telescope or instrument ``breathing.'' This was not a significant issue for the $\alpha$~Cen time series, however, because the E140M observations were carried out exclusively with the high-throughput $0.2^{\prime\prime}{\times}0.2^{\prime\prime}$ photometric aperture, which is large enough to avoid major temporal changes in transmission due to instrumental drifts.  (Changes in the effective throughput can come into play in other circumstances, especially when very narrow slits are used.)  Then, the scaled sub-exposures were weighted in the co-addition by the total of the net counts in each frame, which equivalently is proportional to an effective exposure time.  This processing step is called ``{\small\sf STAGE ZERO}'' in the ASTRAL parlance.

Then, the {\small\sf STAGE ZERO} spectra for the nine visits of, say, A, were combined into an epoch-averaged composite.  Again, cross-correlation of \ion{O}{1} $\lambda$1306 was exploited to align the visit-level spectra in velocity, but in this case the flux densities were given equal weight in the co-addition, without regard to exposure time or total net counts.  Thus, the final co-added spectrum is a true -- albeit sparsely sampled -- epoch average, treating each observation as a separate, independent realization of the stellar FUV energy distribution at the time of observation.  This step is analogous to ``{\small\sf STAGE ONE/TWO}'' of ASTRAL, although with a different weighting scheme (ASTRAL focuses on maximizing S/N rather than achieving a fair temporal average).  During this process, it was noticed that all three \ion{O}{1} triplet components of both stars displayed variable absorption near their line centers, between the different visits.  It is unlikely that the variability has a stellar (or interstellar) origin, but more plausibly is due to changing populations of hot oxygen atoms in the Earth's geocorona, in response to stochastic space weather conditions.

Figure~3 depicts selected wavelength segments from the nine-epoch co-added STIS E140M spectra of $\alpha$~Cen AB, with prominent features identified.  The bottom-most panel, illustrating the near-ultraviolet (NUV) \ion{Mg}{2} doublet, is from archival E230H high-resolution ($R\sim 1\times10^5$) echelle spectra.  The dotted tracing at the bottom edge of A's shading (rarely visible) is the 1\,$\sigma$ photometric error (per resolution element [``resol'']), heavily smoothed; the black dotted curve is the same for B.  The absolute flux density scale is for A; the B fluxes were scaled upward a factor of 3 to compensate for the lower bolometric luminosity of the optically fainter secondary.  On the adjusted flux scale, B is seen to be significantly more active than A (i.e., larger hot-line fluxes in $f/f_{\rm bol}$), as is typical of K dwarfs compared with G dwarfs of the same age, and as was seen in Fig.~1 in terms of $L_{\rm X}/L_{\rm bol}$.  

Note that the FUV continua of the two stars are similar below 1400~\AA, when the factor of 3 bolometric shift is accounted; but the difference in the continua above 1500~\AA, and especially in the near-UV at the \ion{Mg}{2} h and k lines, is much larger.  The FUV continuum is controlled mainly by chromospheric temperatures, which are similar in the two stars, whereas the longer wavelength continuum is purely photospheric and reacts strongly to the $\sim$600~K $T_{\rm eff}$ difference between warmer A and cooler B (e.g., Porto de Mello et 
al.\ 2008).  Incidentally, the third and fourth panels from the bottom, and the bottom-most, correspond to the three spectral windows captured by the solar {\em Interface Region Imaging Spectrometer}\/ (IRIS: De Pontieu et al.\ 2014).

\subsubsection{Emission Line Measurements}

One of the main motivations of the present work was to explore the global plasma dynamics that is imprinted onto the hot-line profiles, in an effort to improve upon the initial study of AB based on {\em HST}\/ Goddard High-Resolution Spectrograph (GHRS)  material (Wood, Linsky, \& Ayres 1997), and later, Pagano et al.\ (2004) with STIS.  The former authors found, in particular, that the \ion{Si}{4} features of both A and B had distinctly non-Gaussian profiles, but could be decomposed straightforwardly into a pair of Gaussian components, one narrow the other broad.  These components also displayed systematic Doppler shifts relative to the stellar rest frame, analogous to redshifts of FUV Transition Zone (TZ: $T\sim 5\times10^{4}$--$5\times10^{5}$~K) lines in the solar context (Brekke et al.\ 1997).  Unfortunately, only the \ion{Si}{4} region, among the three sets of resonance hot lines, was observed in AB by Wood et al., and the S/N and wavelength accuracy of first-generation GHRS, while a great improvement over previous UV instruments, cannot match what now can be achieved with second-generation STIS.  For example, STIS can capture the entire FUV region all at once with E140M, compared with the only tens of \AA\ spectral snapshots of GHRS.  Further, even the medium-resolution echelles of STIS have about twice the spectral resolution of the GHRS first-order dispersers, and excellent wavelength coherence across the whole range, locked in by frequent measurements of the calibration lamps.  

Pagano et al.\ (2004) substantially improved upon the GHRS work by collecting a series of STIS E140H and E230H exposures of $\alpha$~Cen A to cover the full ultraviolet range, from below Ly$\alpha$ up to the atmospheric cutoff ($\sim 3200$~\AA), at high resolution ($R= 1\times10^{5}$ [3~km s$^{-1}$]), focusing mainly on the emission-line-rich FUV region in their published study.  Although the S/N in their spectra was good compared with the previous GHRS measurements, it was somewhat limited by their use of the narrow spectroscopic slit, $0.2^{\prime\prime}{\times}0.09^{\prime\prime}$, and the fact that the FUV, for example, required three separate E140H settings to cover it.  And, unfortunately, only A was observed.  Nevertheless, exploiting the broad spectral grasp afforded by STIS, the authors were able to measure the three important hot-line doublets, in addition to many other relevant species, including key coronal forbidden line [\ion{Fe}{12}] $\lambda$1242.  Like Wood et al.\ (1997), Pagano et al.\ (2004) found that the hot lines were best matched with bimodal Gaussian profiles, and that both components were redshifted relative to narrow chromospheric species (exclusively \ion{Si}{1} emissions in their study).  They reported generally increasing redshifts with increasing formation temperature through \ion{O}{4} ($\sim 1.5{\times}10^{5}$~K), although the hotter \ion{N}{5} and \ion{O}{5} features fell below the upward extrapolation of the trend.  The measured A redshifts ($\sim 5$~km s$^{-1}$) were about half the peak values recorded in disk-center quiet-Sun spectra, as might be anticipated if the (down)flows primarily are radial.  The narrow components of the bimodal fits tended to be more redshifted than the broad components, by $\sim$3 km s$^{-1}$ on average; and the broad components accounted for slightly less than half the total flux, again on average.

The present study steps back from the high resolution of the Pagano et al.\ (2004) program, to make room for observations of both A and B, noting that according to the widths measured by those authors (${\rm FWHM}\sim 50$~km s$^{-1}$), the stellar hot lines would be fully resolved already at medium resolution ($\sim 7$~km s$^{-1}$).  Further, thanks to the joint {\em Chandra/HST}\/ program, not only is there a long FUV time series of AB available to examine possible changes in the narrow/broad decomposition with changing activity levels, but also the different epochs can be combined into a high-S/N average to obtain accurate measurements of faint features that nonetheless are diagnostically valuable.  Because the pointings were conducted at different times of the year, and during different parts of the spacecraft orbit, the telluric and orbital Doppler shifts moved the spectrum onto different sets of pixels, thereby mitigating fixed pattern noise, and allowing the S/N to attain in practice what the photon statistics would predict in principle.  For the single-epoch FUV observation of Pagano et al.\ (2004), the line peak S/N ranged from less than 15 for the faintest of the hot lines, \ion{N}{5} $\lambda$1242, to nearly 50 for the brightest, \ion{C}{4} $\lambda$1548.  Here, the S/N achieved in the epoch-averaged E140M spectrum was 60 for $\lambda$1242 and 200 for $\lambda$1548, equivalent to an effective exposure about 10 times deeper (taking into account the E140H/M resolution difference).

Nevertheless, there are fundamental limits imposed on the quality of kinematic information that can be gleaned from FUV emission line measurements, devolving from: (1) hidden blends; (2) uncertainties in laboratory wavelengths; and (3) measurement schemes (e.g., not making best use of the available information).  The first issue is perhaps the trickiest to deal with, unless there is sufficient context to characterize an interloping, accidental spectral component.  In practice, it is best to simply avoid features suspected of being contaminated by a significant blend (or blends), unless the target feature is, say, a critical part of a density-sensitive line ratio, and there is no good alternative.  An example is the \ion{S}{4}] $\lambda$1404 contamination of density-sensitive \ion{O}{4}] $\lambda$1404, described later.  

The second and third limitations mentioned above --- reference wavelengths and measurement techniques --- are important parts of the analysis, but peripheral to the main results, so were relegated to the Appendices.  The adopted reference wavelengths are listed in Table~3.  These mainly are laboratory values, but include semi-empirical results from measurements of astrophysical sources (a recombination nebula and metal-polluted white dwarfs [WD]) for several of the key hot lines.  Figure~4 summarizes the types of fitting schemes applied to the $\alpha$~Cen datasets.  Tables~4a and 4b list line profile measurements derived from the epoch-averaged spectra of A and B, respectively, including standard deviations of the various quantities over the nine individual epochs to illustrate the magnitude of variability.

\section{ANALYSIS}

\subsection{Cycle Dependence of FUV Fluxes}

Figure~5 illustrates the coronal cycle dependence of FUV lines of various key species having different temperature sensitivities, all based on integrated fluxes (except for \ion{He}{2}, which was taken from a double Gaussian fit to account for a close \ion{Fe}{2} blend).  The multiplet members were added together in the cases of \ion{C}{2} and \ion{C}{4}.  Points represent the line fluxes divided by the average over the nine epochs.  Gray shaded dashed curves are the schematic contemporaneous X-ray cycles (Fig.~1) divided by an average over the same time span as the FUV measurements.  The \ion{He}{2} $\lambda$1640 Balmer-$\alpha$ emission is thought to be formed, at least partly, by a process that is responsive to coronal XUV irradiation (Jordan 1975).

For $\alpha$~Cen A, there is little variation of any of the diagnostic fluxes over the four year period captured by STIS, despite the factor of $\sim$2 increase in the X-ray luminosity, except perhaps for [\ion{Fe}{12}], which, to be sure, should be most sensitive to any changes in coronal conditions.  The [\ion{Fe}{12}] variation of A, however, is compromised by the low S/N of the measurements.  Further, the anomalous velocity behavior of [\ion{Fe}{12}] in A (described later) cautions that the measured feature in the primary might be affected by a blend, such as \ion{S}{1} $\lambda$1241.91 suggested by Pagano et al.\ (2004).  At the same time, the $\alpha$~Cen B FUV fluxes appear to more accurately track the coronal X-ray evolution, particularly during the more recent epochs when B was falling from its cycle peak in 2012.  Note that the response of a feature becomes more pronounced with increasing formation temperature, as has been seen in global comparisons of, say, X-rays to subcoronal and chromospheric tracers in broad samples of cool stars (Ayres et al.\ 1995).

\subsection{Velocity Properties from Single Gaussian Measurements}

Figure~6a is a summary of single Gaussian fits to representative, isolated FUV emission lines from the nine separate epochs, and the epoch average, of $\alpha$~Cen A.  Figure~6b is similar, for $\alpha$~Cen B.  In each panel, black (and blue) dots refer to low-excitation, narrow emissions of chromospheric temperature species, typically neutrals of abundant elements like C and O.  The average velocity of these features was taken as the zero point for the other species.  Dot-dashed horizontal lines indicate $\pm$1\,$\sigma$ dispersions of these measurements about the mean; blue dots mark reference lines that were eliminated from the average based on iterative sigma-clipping (2\,$\sigma$ threshold).  Green dots refer to TZ hot lines, such as \ion{Si}{4}, \ion{C}{4}, and \ion{N}{5}; red dots are for intermediate temperature species like \ion{C}{2} and \ion{Si}{3}, but also including optically thick neutrals such as \ion{H}{1} Ly$\alpha$ and the \ion{O}{1} 1305~\AA\ triplet.  The single orange symbol is for coronal forbidden line [\ion{Fe}{12}] $\lambda$1242.   In most cases, the error flags are smaller than the symbol sizes, owing to the high-S/N FUV profiles recorded in these nearby bright stars.  

The dispersion among the dozen or so velocity reference lines reflects partly the (small) random errors of measurement, but also any systematic effects such as subtle irregularities in the STIS wavelength scales (which are partially mitigated by the post-pipeline distortion correction described earlier) and errors in the laboratory wavelengths.  A few of the reference lines, e.g., $\lambda$1473, display systematic deviations with respect to the mean in many of the individual epochs, as well as in the average.  These would be potential candidates for laboratory wavelength errors.  For B, the dispersion in reference velocities is higher than for A, owing to the larger statistical measurement errors for the generally fainter lines of B (in absolute flux).  Still, the 1\,$\sigma$ level for each individual epoch ($\sim 0.9$~km s$^{-1}$), about 10\% of the instrumental resolution, is comparable to the expected distortion-corrected wavelength scale accuracy (better than 1~km s$^{-1}$ for E140M).  

The small dispersion of the reference lines about the mean trend in general, and the flat behavior over wavelength, is an encouraging indication that the systematic errors are well controlled.  Further, the average low-excitation velocities of (higher-S/N) A in the nine epochs differed from the photospheric radial velocity predicted by the Pourbaix et al.\ (2002) ephemeris by a negligible $+0.1{\pm}0.6$~km s$^{-1}$.  The small standard deviation of the velocities is consistent with the $\sim 1$~km s$^{-1}$ 2\,$\sigma$ centering errors described earlier.  (The agreement was less good for B, however: $+0.5{\pm}1.8$~km s$^{-1}$.)

In fact, the IRIS solar UV imaging spectrometer mentioned earlier utilizes narrow chromospheric emission features to define its dispersion function and wavelength zero point, because the instrument lacks the wavelength calibration lamps that have proved so valuable for STIS (and predecessors).  The STIS measurements here, especially of solar twin A, support the use of the narrow low-excitation chromospheric emissions for these purposes.
 
Note, also, that the hot lines of A (including \ion{O}{5}] $\lambda\lambda$1218,1371, \ion{O}{4}] $\lambda$1401, and \ion{N}{4}] $\lambda$1486) all seem to display a similar, positive (i.e., redshifted) velocity with respect to the low-excitation chromospheric average (with the conspicuous exception of [\ion{Fe}{12}], which shows a blueshift of comparable or larger magnitude).  The hot lines of B are somewhat less redshifted than those of A, but now the [\ion{Fe}{12}] feature exhibits a small redshift, although less than the TZ species.  To be sure, the rest wavelength of the coronal forbidden line is not well enough known to make any definitive statements concerning its Doppler shifts, but the differential behavior between A and B is significant and noteworthy.  In contrast to the nearly constant Doppler shifts here, Pagano et al.\ (2004) found a distinct trend with temperature for their hot-line measurements, albeit with large error bars due partly to lower S/N and partly to disagreements between velocities of doublet members (which should be the same if the components are optically thin).  The difference between the constant Dopper shift with temperature seen here and the trend identified by Pagano et al.\ can be traced mainly to the different reference wavelengths assumed in the independent studies.

\subsection{Collective Hot Line Modeling}

The systematic behavior of the $\sim 10^5$~K TZ lines of both stars provided motivation to consider a more collective approach to empirically characterize their kinematic properties, as illustrated in Figure~7.  The left panels contain, separately for AB, a superposition of the \ion{Si}{4}, \ion{C}{4}, and \ion{N}{5} doublets (six components altogether), interpolated onto the same velocity grid, relative to the individual reference wavelengths, and adjusted to the same integrated flux in the $\pm$65~km s$^{-1}$ interval (marked by vertical dot-dashed lines in Fig.~7) where the features appear to be least affected by extraneous blends.  The spectral traces are from the epoch averages.  Yellow dots represent the mean profile after application of an ``Olympic'' filter (throwing out the highest and lowest flux density in each velocity bin) for $|\upsilon|\lesssim 65$~km s$^{-1}$, and a more aggressive filtering (deleting the 3 highest values) for the flanking $|\upsilon|> 65$~km s$^{-1}$ intervals where the influence of weak emission blends on the faint high-velocity wings of the hot lines is most evident.  The standard error of the mean of the surviving scaled fluxes was taken as an empirical measure of the point-to-point uncertainty.

The filtered hot-line profiles of A and B are illustrated in the right hand panels (larger yellow dots), now on a linear relative flux density scale.  Also shown are bimodal Gaussian fits, explicitly illustrating the E140M $0.2^{\prime\prime}\times0.2^{\prime\prime}$ aperture line-spread function.  The collective approach assumes that all three doublets essentially are carrying the same kinematic information, but the individual profiles might be corrupted to a greater, or lesser, degree by coincidental emission (or possibly also absorption) blends.  The multi-layered filtering attempts to minimize the influence of blends at the hot-line peaks, while still retaining maximum information concerning the average profile there; but also achieve good suppression of emission blends in the far line wings, where the influence of the broad components dominates.  In both sets of panels, thin dashed curves represent single Gaussian fits.  In the linear flux density diagrams (to the right), the single Gaussian models appear to work tolerably well; but in the logarithmic tracings to the left, the far-wing departures of the pure Gaussian models are more conspicuous.  Parameters of the bimodal fits for the epoch-averaged spectra, and standard deviations over the nine individual epochs, are provided in the final rows of Tables~4a and 4b.  Uncertainties on the epoch-averaged profile fits were estimated by the same Monte Carlo approach described in Appendix~B.

\subsection{Summary of Bimodal Hot-Line Profile Fitting}

Figure~8 illustrates the hot-line double Gaussian parameters of $\alpha$~Cen AB, for the individual species and the collectively filtered profiles; over the nine separate epochs and for the epoch-averaged spectra.  The left-most frame compares centroid velocities against the doublet integrated fluxes in each epoch (total \ion{Si}{4}\,+\,\ion{C}{4}\,+\,\ion{N}{5} flux for the collective profiles, multiplied by 1.5 for display purposes), which is a surrogate for activity level.  For both stars, redshifts of the narrow and broad components are similar, about 3--5~km s$^{-1}$; the narrow components systematically are more redshifted in A, while the opposite is true of B.  Also, in both stars, the \ion{Si}{4} shifts tend to be slightly smaller (by $\sim 1$~km s$^{-1}$) than those of \ion{C}{4} and \ion{N}{5}.  The shifts of the collective profiles are in the middle.  There is no obvious change in the narrow and broad component velocities over the range of activity covered by A and B, although admittedly that range is rather compressed for A.  

The middle panel depicts the widths of the narrow and broad components as a function of activity state.  The components of B are slightly narrower than those of A, but in both cases the broad components are about twice as wide as the narrow ones.  Again, no dependence on activity level is obvious.  

The final panel, on the right, displays the relative fluxes, broad divided by total, against activity.  For both stars, the ratios are close to 0.5, reflecting a near equality in the fluxes carried by the narrow and broad components.  

\subsection{Transition Zone Densities from \ion{O}{4}] Line Ratios}

As noted earlier, Flower \& Nussbaumer (1975) proposed that ratios between members of the \ion{O}{4}] intercombination multiplet at 1400~\AA\ are density sensitive and can be used to probe plasma properties at the line formation temperature $\sim 10^5$~K over the range $n_{\rm e}\sim 10^{10}$--$10^{12}$ cm$^{-3}$ thought to be characteristic of solar TZ conditions.  There then ensued a long history of applications of the diagnostic to solar and stellar plasmas (see, e.g., Cook et al.\ 1995, and references therein), which has continued to the present day (e.g., Keenan et al.\ 2009).  

Figure~9 illustrates \ion{O}{4}] ratios derived from the epoch-resolved spectra of AB, and the epoch averages (horizontal shaded bands).  The leftmost panels illustrate $R_{3}= {\lambda}1399/{\lambda}1407$, in the notation of Keenan et al.\ (2002).  This particular ratio should be independent of density, owing to a shared upper level.  The constant value, 1.03 according to the calculations of Keenan et al.\ (2002), is indicated by arrows in the figure.  For both A and B, the observed ratios are close to unity, with small error bars, at least in A, and especially for the epoch-averaged spectrum.  Thus, in subsequent ratios such as $R_{1}= {\lambda}1407/{\lambda}1401$ and $R_{2}= {\lambda}1407/{\lambda}1404$\footnote{Keenan et al.\ (2009) call this ratio $R_{3}$.}, ${\lambda}1407$ was replaced by $0.5\,({\lambda}1399 + {\lambda}1407)$, to boost S/N.  The hybrid ratios are denoted $R\ast$.  

The rightmost panels depict $R_{4}= {\lambda}1397/{\lambda}1404$, also in the nomenclature of Keenan et al.\ (2002).  Like $R_{3}$, this ratio should be independent of density, with a theoretical value of 0.131.  Again, the expectation is met, although the scatter of points, and individual error bars, are larger in keeping with the weakness of the $\lambda$1397 feature.  Note that the key \ion{O}{4}] lines in the Sun --- $\lambda$1399, $\lambda$1404, and $\lambda$1407 --- were challenging to measure even with {\em SoHO}\/ SUMER, and reliable density diagnostic ratios were reported by Keenan et al.\ (2002) only for a sunspot plume, where the TZ emissions are strongly enhanced.

The center two panels of Fig.~9 illustrate the two important density-sensitive \ion{O}{4}] ratios, $R_{1}\ast$ and $R_{2}\ast$, mentioned earlier.  The latter has a $\sim 3\times$ larger grasp (limiting high density value of $R$ divided by limiting low density value of $R$) than the former, although both have similar low and high density limits: $n_{\rm e}\sim 10^{10}$--$10^{12}$ cm$^{-3}$.  Arrows mark the most recent low density limiting values of the ratios reported by Keenan et al.\ (2009; noting that their new $R_{3}$ is the old $R_{2}$).  Both $\alpha$~Cen stars have $R_{1}\ast$ and $R_{2}\ast$ ratios above the low density asymptotic values, especially $R_{2}\ast$ of B.  In each star, the two ratios are jointly consistent with the same density: $10^{9.7}$ cm$^{-3}$ for A, and $10^{10.0}$ cm$^{-3}$ for B.  These values are slightly less than obtained by Keenan et al.\ (2002) for the SUMER sunspot plume from $R_{1}$ ($\sim 0.21{\pm}0.04$), but consistent within the larger error bars of the solar value.  The stellar values suggest that there is not much difference in the TZ conditions of AB, even despite the larger activity levels of B; and little, if any, evolution of those conditions, especially over the rise and fall of B's X-ray cycle.

\section{DISCUSSION}

The similarity of the narrow and broad component fluxes of both $\alpha$~Cen stars is challenging to understand, in some respects, within the context of spatially resolved solar FUV measurements.  A natural analog for the broad components are the so-called TZ ``explosive events,'' which are high-velocity red/blue profile excursions seen in long-slit images of hot lines like \ion{C}{4} (Dere et al.\ 1989).  However, the apparent fractional area covered by such episodes is only a few percent, or so (see Wood et al.\ 1997), contrasted with the $\sim$50\% contribution by the broad components here.  This discrepancy originally was noted by Wood et al.\ (1997), although they obtained a smaller \ion{Si}{4} $\lambda$1393 broad component flux, $\sim 25$\% of total, for a $\sim 50$\% broader and {\em blueshifted}\/ feature of $\alpha$~Cen A, but with the observational caveats mentioned earlier for GHRS.  At the same time, the bimodal widths, shifts, and relative fluxes obtained here for A agree well with the single-epoch STIS values reported by Pagano et al.\ (2004), modulo small differences attributable to the fundamental reference wavelengths.  More recently, Linsky et al.\ (2012) found that a nearly equal split between the narrow and broad component fluxes extends even to extremely active young solar analogs, like 50~Myr EK~Draconis (see, also, Ayres \& France 2010).

The apparent dichotomy highlighted by Wood et al.\ and Pagano et al.\ was not lost on Peter (2006), who carried out, in many respects, an heroic data analysis effort to try to pin down the spatial origin of the narrow and broad components in solar \ion{C}{4} $\lambda$1548, exploiting a unique all-sun scan of that feature by {\em SoHO}\,/SUMER.  Peter constructed a disk-average spectrum of $\lambda$1548 by fitting a single Gaussian profile to each spatial point spectrum along the SUMER slit, then averaging these over all the slit positions that constituted the full-sun scan.  Peter notes that this strategy specifically suppresses the faint broad wings occasionally seen from localized explosive phenomena, so they are in effect excluded from the final cumulative line shape.  He then modeled the disk-average profile using the bimodal Gaussian approach.  He obtained a narrow component of ${\rm FWHM}\sim 42$~km s$^{-1}$ and a broad component about twice as wide ($\sim 70$~km s$^{-1}$), similar to the epoch-averaged values here for $\alpha$~Cen A (41~km s$^{-1}$ and 80~km s$^{-1}$, respectively) and to the single-epoch measurements of $\lambda$1548 by Pagano et al.\ (43~km s$^{-1}$ and 79~km s$^{-1}$, respectively).  He also found nearly equal flux contributions by the two components, slightly favoring the narrow one.  Here, for nominal solar-twin $\alpha$~Cen A, the components also are nearly equal, although now the broad component is slightly favored.  The striking difference between Peter's full-disk solar \ion{C}{4} profile and that for A here, is that he obtained only a small redshift for the narrow component ($\sim +1$~km s$^{-1}$), but a larger redshift for the broad component ($\sim +6$~km s$^{-1}$); whereas here for A the narrow and broad components have similar redshifts ($\sim +5$~km s$^{-1}$), slightly larger for the narrow component.

The story might end there, with, say, a speculative discussion of why the SUMER profile spatial averaging technique might be biased in one way or another, and how the observed profile of nominal solar twin $\alpha$~Cen A, which represents a completely unbiased disk average, must be a better representation of what a true solar disk average should look like.  There is, however, one additional point to be considered, namely $\alpha$~Cen B.  Curiously, the redshift and FWHM of the \ion{C}{4} broad component (+5.6, 69~km s$^{-1}$) are nearly identical to what Peter (2006) found for the SUMER broad component, and like the solar \ion{C}{4} profile, the B narrow component is much less redshifted than the broad one, and is slightly larger in flux contribution (although about 20\% narrower in FWHM).  These similarities might, of course, be coincidental.  However, Linsky et al.\ (2012) also identified a trend among their active G stars of a progressively less redshifted narrow component compared to the broad component, as a function of increasing activity.  In that context, it is worth recalling that in Fig.~1 the Sun falls midway between A and B in in terms of peak activity level, $(L_{\rm X}/L_{\rm bol})_{\rm max}$.  Thus, ironically, it might be B that is the better solar analog, at least in terms of TZ activity.  It also is worth reiterating Peter's conclusion that the narrow/broad partition of the TZ line profiles, especially the broad components, are more representative of long-lived surface structures, like the chromospheric network bright points, rather than a direct consequence of transient explosive heating events (which must be present in disk-average profiles, but at a very low, inconspicuous, level).

Finally, the fact that the narrow/broad flux partition does not seem to depend on activity level, either during the cycle evolution of B, or between A and B, nor do the average densities seem to evolve either, suggests that the dominant surface sources of the FUV fluxes are relatively unchanged during the cycle, aside perhaps for the fractional area covered.  Returning to Peter's conclusion concerning the importance of the fine scale magnetic elements --- \ion{Ca}{2} K bright points --- in the quiet-Sun supergranulation pattern, one also should include consideration of the similarly fine scale magnetic elements in large-scale plage regions surrounding sunspots.  The main qualitative difference between the two types of bright points is the higher surface covering fraction (points per unit area) in the plage, something like 5 times more than in the supergranulation network.  However, the supergranulation is ubiquitous on the solar disk, and more-or-less independent of cycle phase; whereas the plage component occupies a much smaller portion of the disk (albeit with its higher internal filling factor), but that portion varies strongly with the sunspot number, and therefore also the activity cycle.  Perhaps, then, one can think of a ``quantum'' of surface activity, sprinkled lightly along the supergranulation subduction lanes, but more heavily in magnetic active regions forming the signature extensive plage.

One still must explain why the kinematic behavior, in particular the component redshifts, are ostensibly different between A and B.  In this regard, it would be worth repeating the SUMER full-sun spatial scan experiment, but now using the new IRIS solar imaging spectrometer, which in many technical respects is superior to the previous FUV workhorse on {\em SoHO}.

The joint {\em Chandra/HST}\/ campaign on $\alpha$~Cen AB currently is scheduled to extend through 2017.  Barring instrumental issues, this period likely will see B falling to the next minimum of its cycle, while A is climbing toward a possible peak of its own.  This crossroads will provide significant leverage on the cycle dependence of TZ conditions on AB, a key piece in the unfolding puzzle of high-energy surface activity on sun-like stars.  Time will tell.   

\clearpage
\acknowledgments

This work was supported by grants GO-12374 and GO-12758 (which includes programs GO-13060 and GO-13465) from the Space Telescope Science Institute, based on observations from {\em Hubble Space Telescope}\/ collected at STScI, operated by the Associated Universities for Research in Astronomy, under contract to NASA.  Additional support was by GO1-12014X and GO2-13018X from the Smithsonian Astrophysical Observatory, based on observations from {\em Chandra}\/ X-ray Observatory collected and processed at the CXO Center, operated for NASA by SAO.  This research also made use of public atomic physics databases organized under GENIE (https://www-amdis.iaea.org/GENIE/).  

Facilities: \facility{HST(STIS)}, \facility{Chandra(HRC)}

\clearpage
\appendix

\section{Reference Wavelengths}

There are a variety of popular tabulations of atomic physics parameters for spectral lines, especially the laboratory wavelengths.  Perhaps the most fundamental of these is the National Institutes of Standards and Technology (NIST) Atomic Spectra Database (ASD: Kramida et al.\ 2013)\footnote{see: http://www.physics.nist.gov/PhysRefData/ASD/lines\_form.html}.  Two types of wavelengths are provided: ``Observed'' and ``Ritz.''  The latter are calculated from energy level differences, and in many cases should be preferred over observed values, especially outdated measurements.  In particular, a given spectral transition, say in the ultraviolet, might be difficult to produce and record directly in the laboratory, whereas the associated energy levels might be more accessible via optical or infrared transitions that arise from, or terminate, at those levels.  On the other hand, there could be specialized cases for which a more recent laboratory wavelength measurement might be preferred over an older Ritz value based on less up-to-date energy levels.

There are three general areas where accurate wavelengths are needed: (1) low-excitation, narrow transitions that are used to validate the zero point of the STIS wavelength scales, in the face of small target centering errors and instrumental drifts that can introduce spurious Doppler shifts; (2) the hot lines, themselves, which carry the kinematic signatures of heating and cooling processes in the low corona; and (3) density-sensitive multiplets, where precise relative wavelengths for the transition array can be exploited in constrained multi-line fitting to ensure accurate fluxes, especially for the fainter members.

A comparison of low-excitation line wavelengths, e.g., for \ion{O}{1}, \ion{C}{1}, \ion{S}{1}, and \ion{Fe}{2}, among the various atomic databases generally finds good agreement with the NIST Ritz values at the ${\pm}1$~m\AA\ level, which corresponds to about 0.2~km s$^{-1}$ at 1400~\AA.  This is sufficient accuracy given the residual distortion errors in the STIS E140M wavelengths, perhaps several times that level (see Ayres 2008).  Thus, the NIST Ritz values were adopted for this type of transition.

Among the hot lines, the \ion{Si}{4} and \ion{N}{5} resonance doublets have the same values for Observed and Ritz wavelengths.  For \ion{C}{4} $\lambda$1548, however, O--R differs by +15~m\AA, and for $\lambda$1550 by +2~m\AA.  The latter difference is perhaps tolerable, but the former corresponds to about 3~km s$^{-1}$, which is comparable to the magnitude of the hot-line Doppler shifts reported for AB by Wood et al.\ (1997) and for A by Pagano et al.\ (2004).  In fact, the NIST wavelengths for these key high-ion resonance transitions date back a half century or more.  Fortunately, there has been significant progress in recent years in the development of vacuum ultraviolet (VUV) Fourier transform spectroscopy (FTS) together with high-current discharge sources capable of producing the relevant ions.  Griesmann \& Kling (2000) have reported measurements of the \ion{Si}{4} and \ion{C}{4} resonance doublets by that technique, obtaining especially high accuracy for the former, and much improved accuracy for the latter (C$^{3+}$ is difficult to produce in the Penning discharge, so the S/N of the measurements was lower).  Unfortunately, the \ion{N}{5} resonance lines fall below the shortwavelength cutoff of the calcium fluoride optics of the VUV-FTS.  The Griesmann \& Kling wavelengths form the backbone of an empirical bootstrap approach to refine the \ion{N}{5} (and \ion{C}{4} $\lambda$1550) reference wavelengths, described shortly.

The \ion{O}{4}] semi-forbidden multiplet near 1400~\AA\ is another special case.  In fact, NIST ASD does not list this multiplet, although wavelengths can be derived from the associated {O}$^{3+}$ energy levels.  An empirical value for the strongest component, $\lambda$1401, obtained by Harper et al.\ (1999) from careful measurements of GHRS FUV spectra of the recombination nebula of the RR~Telescopii symbiotic binary, is 11~m\AA\ larger than Kurucz's CD-ROM~23\footnote{http://www.cfa.harvard.edu/amp/ampdata/kurucz23/sekur.html} entry (latter is 1401.157~\AA, same as derived from the ASD energy levels); these two values bracket the wavelength given by The Atomic Line List v2.04\footnote{http://www.pa.uky.edu/$\sim$peter/atomic/} (1401.164~\AA), which tabulates a slightly different energy for the $^{2}$P$^{\rm o}_{{3}/{2}}$ lower state of the transition.  More recently, Young et al.\ (2011) carried out an analysis similar to Harper et al.\ (1999) of forbidden and intercombination wavelengths in RR~Tel, but now using high-quality STIS spectra tied into high-resolution optical material from the European Southern Observatory.  They obtained a wavelength for $\lambda$1401 identical to the NIST energy level value.  The Young et al.\ wavelength for \ion{O}{4}] $\lambda$1401 was adopted here.

What is more important than the absolute wavelength of $\lambda$1401, however, is the array of relative values between $\lambda$1401 and the weaker members of the multiplet, at least for the purposes of the constrained fitting described later.  Also, the key density-sensitive member $\lambda$1404 is blended with a faint \ion{S}{4}] line, which too is part of an intercombination multiplet in that region (e.g., Keenan et al.\ 2002).  The \ion{S}{4}] $\lambda$1404 flux can be estimated from nearby \ion{S}{4}] $\lambda$1406, so good wavelengths relative to \ion{O}{4}] $\lambda$1401 are important for these lines as well.  

The issues of the hot-line absolute wavelengths, and the intercombination multiplet relative wavelengths, was addressed here using the same semi-empirical approach exploited by Harper et al.\ (1999) and Young et al.\ (2011; see also Keenan et al. 2002), namely to measure the relevant spectral features in astrophysical, rather than laboratory, sources.  The clear benefit of the approach is that many cosmic environments can produce strong emissions (or absorptions) of the target species, whereas especially the high ion state lines are challenging to populate in normal laboratory discharge lamps.  The downside of the semi-empirical method is that astrophysical sources capable of exciting high-energy transitions usually exhibit significant internal dynamics, which could foil an attempt to precisely measure relative wavelengths, especially between different ion states (e.g., \ion{N}{5} vs.\ \ion{Si}{4}) that might arise in kinematically distinct regions of a complex environment.  Consequently, care in the source selection is essential.

For this purpose, again RR~Tel was consulted for the \ion{O}{4}] and \ion{S}{4}] multiplets.  These features are bright in the RR~Tel nebular emission spectrum and easily measured.  The difference here is that the STIS FUV spectrum of RR~Tel (dataset o5eh01010: a single E140M-1425 2.4~ks exposure through the 0.2$\times$0.2 aperture in 2000 October) was processed through the same ASTRAL protocols as the $\alpha$~Cen echellegrams already described; to mitigate issues with wavelength distortions that were of some concern to the Young et al.\ project and an essential point of focus for a wavelength study.  

The \ion{O}{4}] and \ion{S}{4}] transitions in the 1400~\AA\ region of the RR~Tel spectrum have identical shapes characterized by a peaked, but asymmetric, profile with a noticeably extended red wing.  As in Young et al.\ (2011), the features individually were modeled by a pair of blended Gaussians, although unlike Young et al.\ there was no constraint placed on the relative widths of the two components.  Only the stronger, blue component of the fits was considered subsequently.  Young et al.\ identify this as the nebular feature.  A velocity shift of the measured $\lambda$1401 transition (strongest of the array) was determined relative to the adopted rest wavelength (of Young et al.\ 2011) and applied to the other observed wavelengths of the multiplet members to establish their rest wavelengths.  The values so obtained are reported in Table~3.  The average difference relative to the Young et al.\ (2011) wavelengths is +0.6${\pm}$0.9~km s$^{-1}$, where the error is the standard deviation over the 7 measurements.  The difference is significant at about the 2\,$\sigma$ level with respect to the standard error of the mean, but is small in any event.

As for the hot lines, the features in RR~Tel have stronger contributions from the red component, which appear to increase in step with formation temperature, in the sense that \ion{Si}{4} is the narrowest while \ion{N}{5} is the broadest.  This complication renders meaningful measurements of the relative line positions challenging and uncertain.  Thus, a different type of source was considered for the hot lines, namely metal-polluted DA/DO white dwarfs, specifically the STIS calibration targets G191B2B and BD+28$\degr$4211.  Each of these objects has a very extensive collection of STIS echelle observations, taken during the initial operational period (1997--2004) as well as post-SM4 (2009--present).  

Thanks to G191B2B's role as prime calibrator for the echelle blaze sensitivity curves, it has the unique distinction of having been observed in all 44 STIS echelle settings (1 E140M, 11 E140H in the FUV).  There were two such whole-echelle campaigns: one during the initial phase of STIS operations covering the period 1998--2001, the other shortly after the STIS repair, during a six-week interval at the end of 2009 and beginning of 2010.  The overlapping H-mode observations can be merged to produce a single spectrum of high-S/N, through suppression of fixed pattern noise, and excellent wavelength coherence, by averaging over slight residual wavelength distortions.  

Thanks to BD+28$\degr$4211's role as prime calibrator for sensitivity degradation monitoring, it has an equally unique extensive time series of E140M-1425 and E140H-1416 exposures, covering both of the STIS operational periods.  These time series can be averaged to enhance mainly the S/N, since the changing heliocentric and spacecraft velocities move the echellegrams onto different pixels in each epoch.  There is an additional, but smaller, mitigation of fixed pattern noise through the few-pixel randomness of the $y$-location of the echelle format due to the grating positioning mechanism.

All the relevant echelle observations of each calibration WD were combined using again the ASTRAL protocols, to yield complete FUV tracings of G191B2B in medium and high dispersion, and of BD+28$\degr$4211 in medium resolution for the whole range, and in high resolution for the more restricted E140H-1416 interval (1321--1512~\AA).  S/N in the BD+28$\degr$4211 M spectrum was more than 400 (per resol) at 1240~\AA\ (\ion{N}{5}) and 1400~\AA\ (\ion{Si}{4}), and about 300 at 1550~\AA\ (\ion{C}{4}).  The equivalent values for G191B2B M were 200 and 150, respectively; and for H, 150 and 100.  

The three sets of hot lines then were measured in the coadded spectra of the two WDs, utilizing blended Gaussians if indicated: both stars display broader photospheric absorptions of the hot lines, and for \ion{Si}{4} in BD+28$\degr$4211 and \ion{C}{4} in both stars blueward displaced narrow components that are thought to arise from photoionization of the local interstellar gas by the hot WD (e.g., Dickinson et al.\ 2012).  The component blending for \ion{C}{4} is worse in G191B2B than in BD+28$\degr$4211.  

The most reliable properties of the hot-line wavelengths that can be derived from these high-quality spectra are the doublet separations, because there is less concern that the apparent features might be coming from somewhat different environments (a caution that applies to comparisons of different species).  Nevertheless, one must be wary of, say, unresolved ISM components that might have different strengths for the doublet members owing to the influence of the factor of two difference in the oscillator strengths on saturation of the profiles.  This effect is seen, for example, in the \ion{C}{4} ISM absorptions of G191B2B in the H spectrum: the stronger 1548~\AA\ line is flat-bottomed and reaches zero flux density, while the weaker 1550~\AA\ line is sharper at the bottom and does not reach zero.  

Considering all the \ion{Si}{4} measurements (broad and narrow components in BD+28$\degr$4211 M and H[1416]; broad components in G191B2B; six $\lambda$1402--$\lambda$1393 differences altogether) the average doublet separation was 9.012${\pm}$0.002~(0.001)~\AA, where the leading uncertainty is the standard deviation, and the trailing parenthetical value is the standard error of the mean.  This compares favorably with the laboratory VUV-FTS value of 9.013~\AA, with zero uncertainty for the decimal places reported.  The \ion{N}{5} doublet separation, in a similar way but now only the broader WD features for G191B2B H and M and BD+28$\degr$4211 M (three $\lambda$1242--$\lambda$1238 differences altogether), was 3.990$\pm$0.002~(0.001).  This contrasts with the NIST Ritz value of 3.983~\AA, which is almost 2~km s$^{-1}$ smaller.  Finally, the average \ion{C}{4} doublet separation, $\lambda$1550--$\lambda$1548, considering the broad and narrow components of BD+28$\degr$4211 M and G191B2B H, but only the narrow component of G191B2B M, was 2.567$\pm$0.004~(0.002)~\AA\ for the five independent measurements.\footnote{The G191B2B broad component is more blended at M resolution, and accordingly had a $\sim$4 times larger measurement error than the others, and was excluded on that basis.  Including that value in the average would lower it to 2.564$\pm$0.007~(0.003)~\AA.}  This is about 2~km s$^{-1}$ smaller than the VUV-FTS value of 2.577$\pm$0.002, and inconsistent with it at the 5\,$\sigma$ level (with respect to the s.e.\ of the empirical measurements).

The discordance between the empirical and laboratory values for the \ion{C}{4} doublet separation is troubling, because, again, differences of a few km s$^{-1}$ are comparable to the size of the hot-line Doppler shifts previously reported for the $\alpha$~Cen stars.  There are two obvious systematic effects to consider on the empirical side: (1) the source lineshapes have internal complexity that affects the differential measurement in some way; and (2) possible distortions of the STIS wavelength solution over scales of several \AA.  

The first concern was minimized by considering the two different WDs (and their different accompanying photoionization regions), as well as the contrasting spectral resolutions (at least for G191B2B).  The second concern was tested according to an approach outlined by Ayres (2008), whereby one runs deep exposures of the STIS wavecal lamps through the normal pipeline as if they were observations of an external target, and then post-processes them using the ASTRAL protocols, as for the other stars discussed here.  Comparing the empirical positions of the lamp lines with the laboratory values then can reveal any significant deviations with wavelength, as a gauge of the coherency of the fully processed wavelength scales.  To that end, all of the recent (post-SM4) deep ($t_{\rm exp}\ge 266$~s) lamp exposures taken with E140M, or the various E140H settings, were de-archived and processed through {\sf calstis} as science images (using the lamp exposure as its own reference wavecal to set the zero-point offset), then combined, again using the same ASTRAL coaddition protocols as applied to the WDs, to yield high-S/N H and M full-coverage FUV tracings.  As anticipated from the previous experiments of this type (Ayres 2008, 2010), there were no significant deviations of the lamp lines from their laboratory wavelengths over the FUV H and M intervals.  The 1\,$\sigma$ dispersion of the measurements was 0.2~km s$^{-1}$ for H (290 lines), at the discretization level for the reported wavelengths; and slightly larger, 0.35~km s$^{-1}$ (310 lines) for M, but still equivalent to only 1/20-th of a resolution element.  

Based on these findings, the empirical \ion{C}{4} doublet separation was adopted here, but to hedge bets somewhat, the absolute laboratory (VUV-FTS) wavelength of the stronger 1548~\AA\ line was chosen.  In fact, the observed displacement between $\lambda$1548 and \ion{Si}{4} $\lambda$1393 was 154.443$\pm$0.000~\AA\ in RR~Tel compared with the VUV-FTS value of 154.444$\pm$0.001~\AA, although with the caveats concerning the non-self-similar profiles of the hot lines in that object; and 154.437$\pm$0.045~(0.018)~\AA\ for the 6 combinations in G191B2B H and M and BD+28$\degr$4211 M.  This provides some confidence in the $\lambda$1548 laboratory wavelength, or at least no strong reason to prefer another value.  The $\lambda$1550 \ion{C}{4} component then was obtained from the $\lambda$1548 laboratory wavelength by adding the empirical doublet separation.  The resulting values, perhaps coincidentally, are close to the NIST ``Observed'' entries.

Finally, the \ion{N}{5} $\lambda$1238 wavelength was determined relative to the highest quality VUV-FTS line, \ion{Si}{4} $\lambda$1393.  The displacements, $\lambda$1393--$\lambda$1238, between the WD photospheric features in G191B2B M and BD+28$\degr$4211 M are very similar, 154.950$\pm$0.001~\AA.  Consideration of additional WD spectra for this key displacement, especially WD1738+669, and including the somewhat smaller G191B2B H value, suggests a larger standard deviation of $\pm$0.005~(0.003)~\AA.  That, of course, is exclusive of any systematic errors devolving from the comparison of WD photospheric lines of significantly different excitation.  In this regard, it also should be mentioned that Stark shifts are negligible for the hot lines because they form in atmospheric layers where the electron densities are at most a few$\times 10^{15}$ cm$^{-3}$ in both WDs (e.g., Rauch et al. 2013: G191B2B; Latour et al. 2013: BD+28$\degr$4211), well below the $\sim 10^{17}$ cm$^{-3}$ where the pressure shifts become appreciable (i.e., $> 1$~m\AA)\footnote{see: http://stark-b.obspm.fr}.  Also, the large gravitational redshifts of these compact objects affects all the stellar hot lines equally.  The final adopted reference wavelengths are listed in Table~3.

\section{Line Profile Modeling}

Several different measurement schemes were applied to the $\alpha$~Cen FUV spectra, for the nine individual epochs, and the epoch-averaged tracings.  These strategies are illustrated schematically in Figure~4.  For this example, the underlying spectrum is the epoch-averaged E140M of $\alpha$~Cen A.  Note the several decade logarithmic flux density scale, to simultaneously capture the bright peaks of the strong \ion{Si}{4} doublet, while also highlighting faint spectral structure near the continuum level, such as weak emission blends (e.g., \ion{Ni}{2} in the blue wing of $\lambda$1393) and the semi-forbidden \ion{O}{4}] multiplet, a key plasma density diagnostic for the subcoronal temperature range (Flower \& Nussbaumer 1975).  

The top panel illustrates an integrated flux measurement, corresponding to a wavelength bandpass encompassing the target line, although perhaps excluding obvious blends (e.g., in \ion{Si}{4} $\lambda$1393).

The middle panel is the result of fitting single Gaussians to selected wavelengths in the features individually, again avoiding, to the extent possible, parts of the profiles affected by blends.  Here, the \ion{O}{4}] $\lambda$1401 feature, strongest component of the semi-forbidden multiplet, was measured in addition to the two bright \ion{Si}{4} features.  

The bottom panel depicts, for the \ion{Si}{4} doublet, a constrained bimodal Gaussian fit in which the $\lambda$1393 and $\lambda$1402 spectral profiles were modeled simultaneously, assuming that the widths of the two narrow components were the same, and similarly for the two broad components; the centroid shifts were the same for the two narrow components, relative to the reference wavelengths, and separately for the two broad components (i.e., possibly different shifts between the two sets); and that the narrow/broad component flux ratios were the same in each doublet member, and only the total doublet flux ratio $\lambda$1402/$\lambda$1393 ($\sim$0.5 in the optically-thin limit) was allowed to float.  In addition, a linear, possibly sloping, continuum level was fitted as part of the procedure; as was the wavelength difference between the two lines, to account for possible laboratory discrepancies and/or subtle lingering distortions in the STIS wavelengths.  

As noted in previous work (e.g., Wood et al.\ 1997), the bimodal Gaussian model more closely reproduces the shapes of the hot lines than a pure Gaussian.  For lower S/N features, like the weaker \ion{O}{4}] multiplet components, however, the benefits of the bimodal fit are not as noticeable.  It should be emphasized that the multi-component approach does not in itself have any physical significance, but simply is a convenient way to more accurately represent the apparent internal complexity of the lineshapes.  Possible physical insight could be obtained, nevertheless, through comparisons with spatially resolved solar hot-line profiles, as described in \S{4}.  

Finally, also illustrated in the lower panel, the \ion{O}{4}] multiplet was modeled by a set of single Gaussian profiles (seven altogether: five \ion{O}{4}] plus two \ion{S}{4}]), constraining the widths and centroid shifts of all the components to be the same (i.e., keeping the wavelength differences between the components at the reference values, but allowing a collective Doppler shift for the whole group), and fixing the relative strength of the \ion{S}{4}] blend with \ion{O}{4}] $\lambda$1404 at 30\% the intensity of \ion{S}{4}] $\lambda$1406 (also one of the components fitted), as indicated by plasma emission models (Astrophysical Plasma Emission Database [APED] 2.0.2: Foster et al.\ 2012).  Again, a linear, possibly sloping, background was included in the global fit.  

The various constraints adopted for the \ion{Si}{4} and \ion{O}{4}]\,+\,\ion{S}{4}] modeling scenarios are valid as long as all the components are optically thin, a likely situation for these relatively low-activity stars.  In all cases, the influence of the E140M line-spread function ($\sim 7$~km s$^{-1}$ FWHM) was included.  The fitting procedure was fully automated, driven by scripts developed in a ``training session'' to pin down the various integration and fitting bandpasses for each feature, or group of features.  The same script was applied to both A and B, given the close similarity of their spectra.  Tables~4a and 4b list the full set of measurements for the epoch-averaged FUV spectra of A and B, respectively, including the type of modeling scheme, uncertainties in the fits determined by a Monte Carlo procedure, and dispersions of the fitted parameters over the nine epochs of post-2010 STIS pointings.  The Monte Carlo error approach, itself, began with a reference profile (the initial, possibly multiple,  Gaussian fit), then perturbed it with a random realization of the smoothed photometric noise level, and refitted the noise-distorted feature.  Dispersions of the lineshape parameters accumulated over many such trials were taken to be a fair measure of the modeling errors.

\clearpage
\begin{figure}
\figurenum{1}
\vskip  -10mm
\hskip  -5mm
\includegraphics[bb=95 95 512 679, scale=0.85,angle=90]{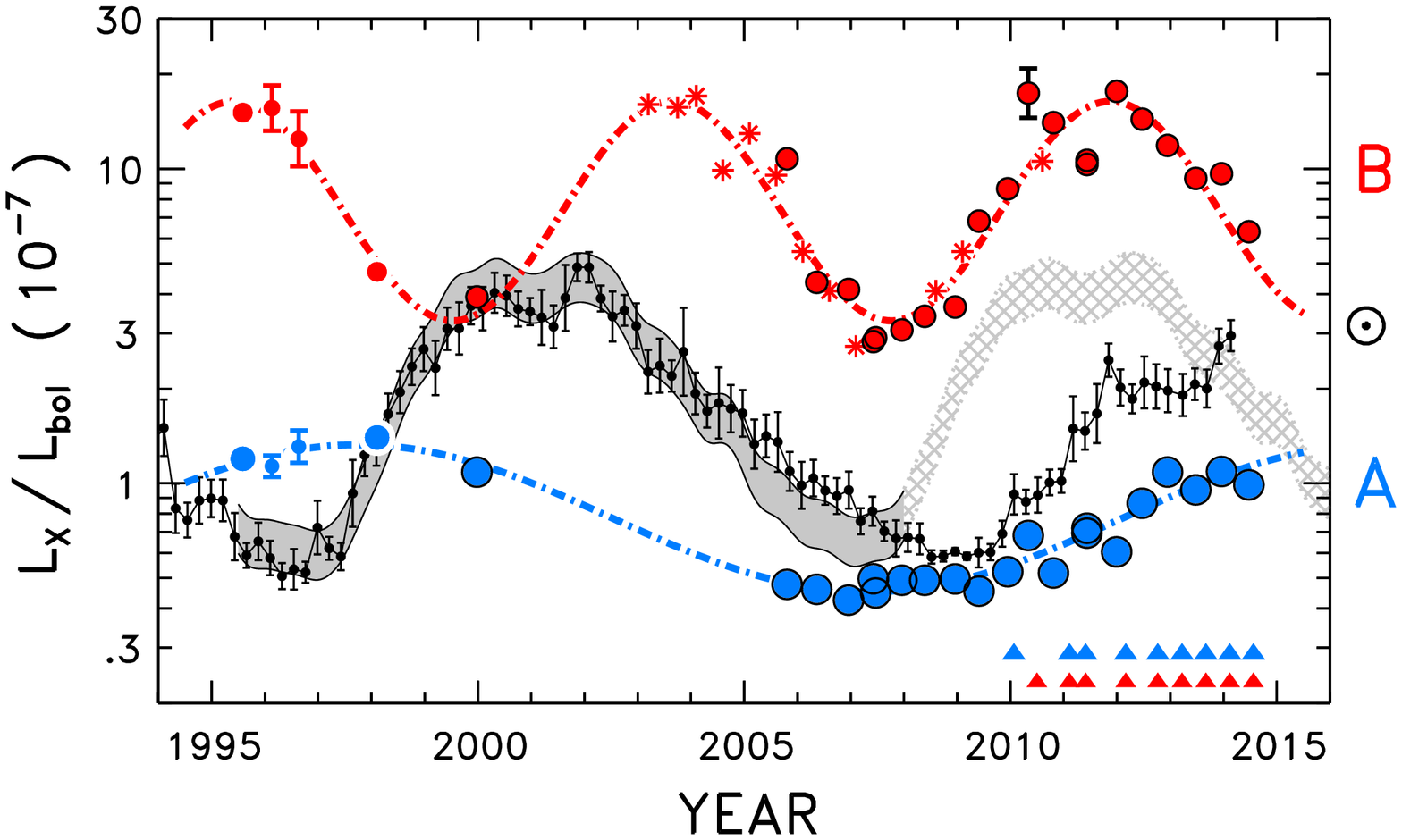} 
\vskip 0mm
\figcaption[]{Coronal X-ray-to-bolometric luminosity ratios of the Sun (black dots and error bars), $\alpha$~Cen A (blue), and B (red).  (Dividing by $L_{\rm bol}$ mitigates the bias introduced by the different stellar sizes.)  Uncertainties of the AB values usually are smaller than the symbol sizes.  Pre-2000 values are from {\em ROSAT}\,; post-2000 from {\em Chandra}\/ (dots) and {\em XMM-Newton}\/ (asterisks: B only).  Shaded background for Sun is schematic three-cycle average.  Triangles mark epochs of post-SM4 STIS pointings.}
\end{figure}

\clearpage
\begin{figure}
\figurenum{2}
\vskip -10mm
\hskip -19.5mm
\includegraphics[bb= 43 70 594 721,scale=0.90,angle=90]{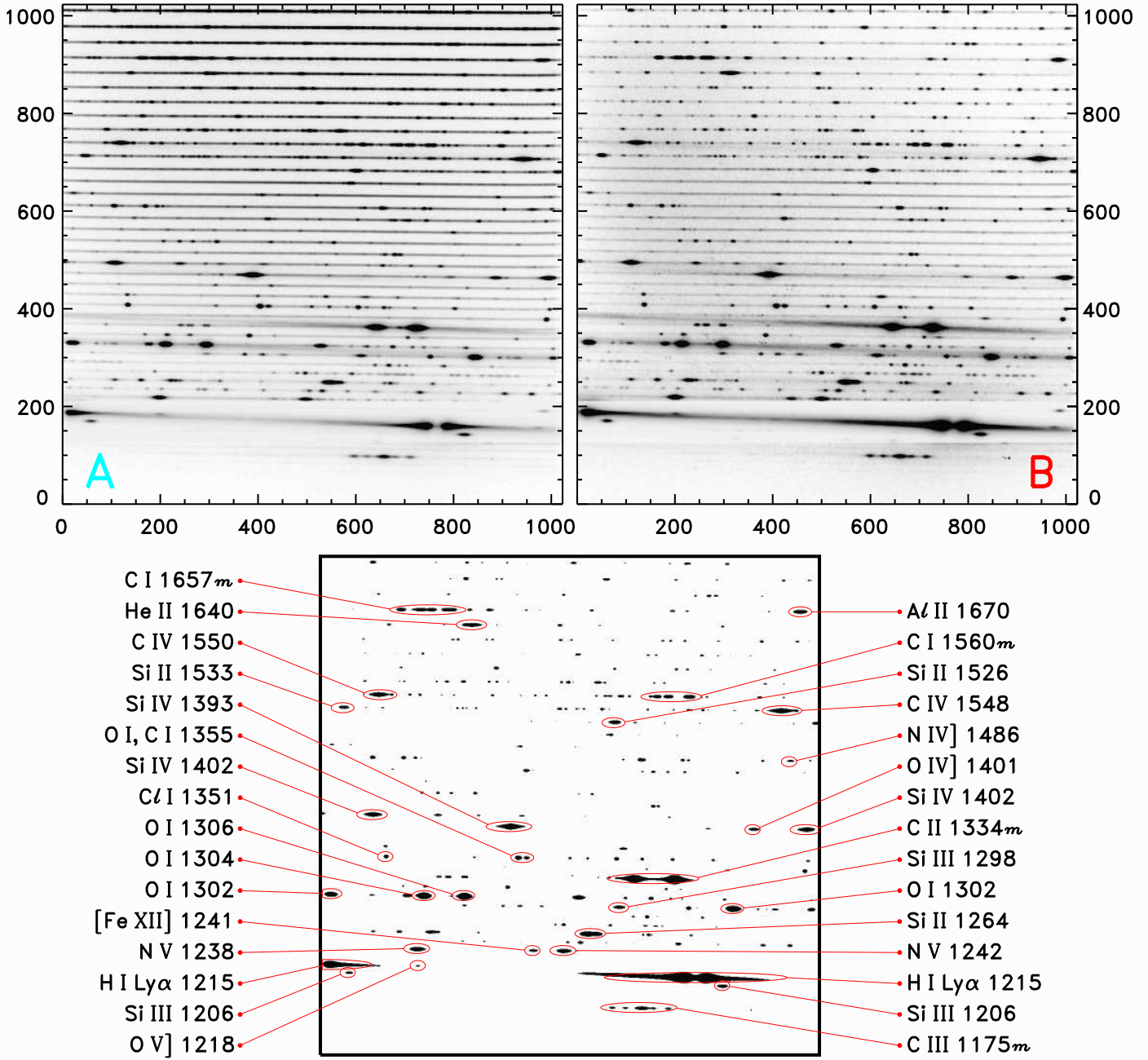} 
\vskip 5mm
\figcaption[]{Negative image (dark\,=\,high signal) E140M-1425 echellegrams of $\alpha$~Cen A (upper left) and B (upper right), summed over the nine epochs.  A map of prominent features is at lower center.  Suffix ``{\em m}'' indicates a multi-transition multiplet.}
\end{figure}

\clearpage
\begin{figure}
\figurenum{3}
\vskip -8mm
\hskip -2mm
\includegraphics[bb= 28 7 549 771,scale=0.875,angle=0]{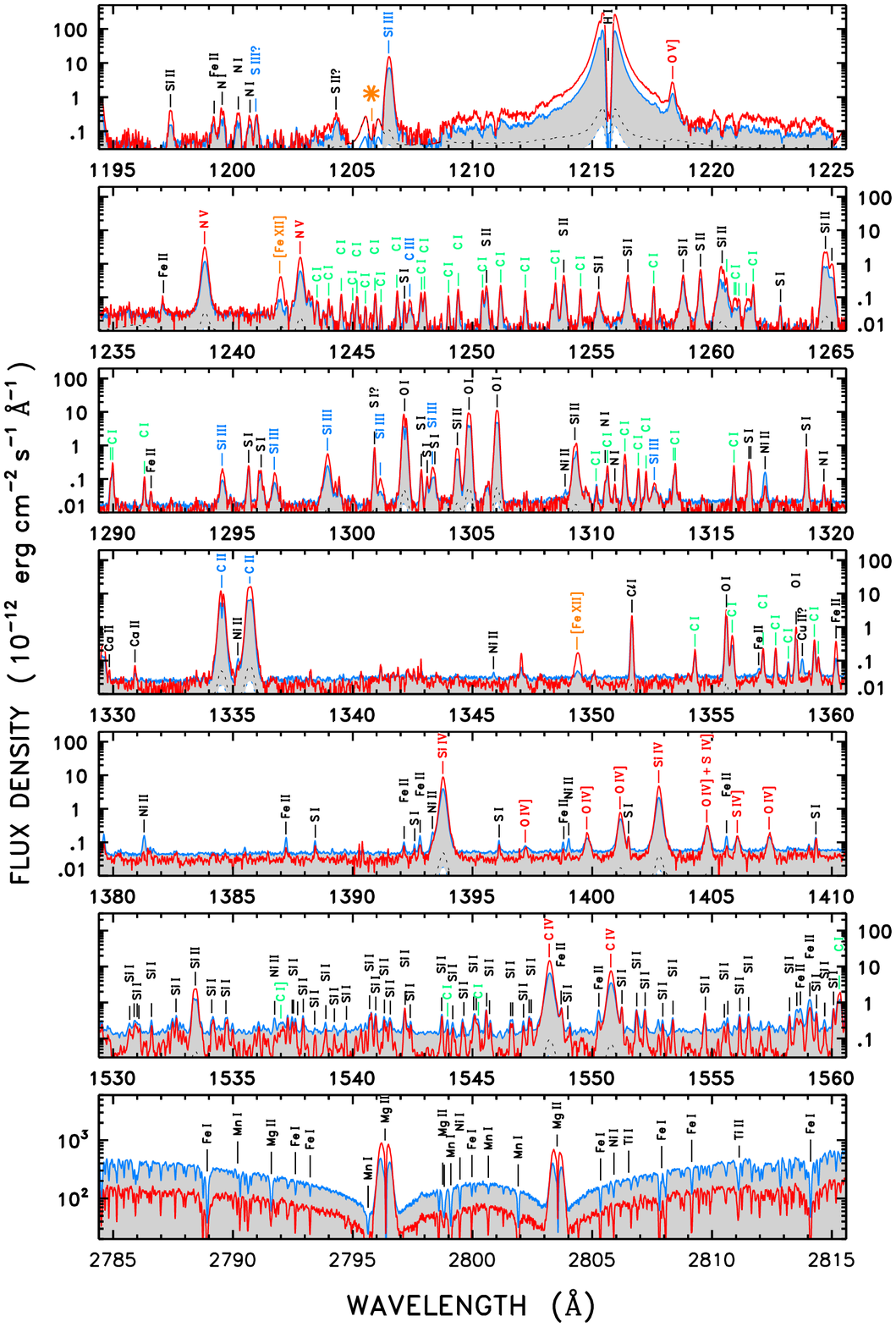} 
\vskip -7mm
\figcaption[]{}
\end{figure}

\clearpage
\begin{figure}
\figurenum{3}
\figcaption[]{Selected wavelength segments from epoch-averaged E140M spectra of $\alpha$~Cen A (grey shaded, blue outline) and B (red).  Black dotted curve is 1\,$\sigma$ photometric error (per resol) for B; 1\,$\sigma$ level for A (lower boundary of shading) mostly is off-scale at bottom of each panel.  In top-most panel, the orange asterisk to left of \ion{Si}{3} $\lambda$1206 marks a region affected by scattered Ly$\alpha$ light bleeding down from one echelle order up.  Colors of the line designations have the following meanings: green for \ion{C}{1}; black for other low-excitation chromospheric features; red for the TZ ``hot lines;'' blue for intermediate-excitation species like \ion{C}{2} and \ion{Si}{3}; and orange for coronal forbidden lines of \ion{Fe}{12} (1242~\AA, 1349~\AA), enhanced in more active B.}
\end{figure}

\clearpage
\begin{figure}
\figurenum{4}
\vskip -20mm
\hskip -21mm
\includegraphics[bb= 52 77 556 714,scale=0.85,angle=90]{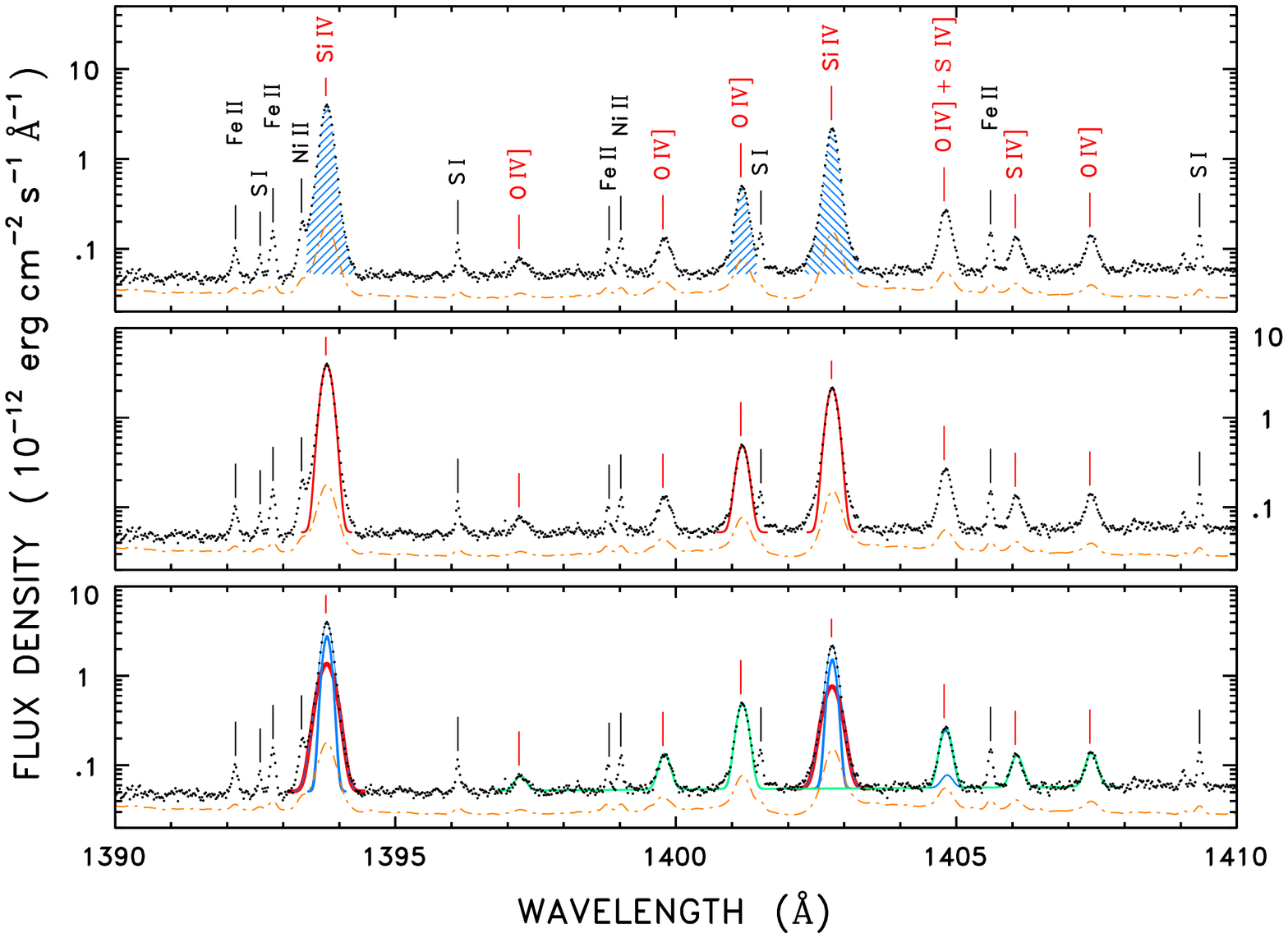} 
\vskip 0mm
\figcaption[]{Examples of measurement schemes applied to the $\alpha$~Cen AB FUV spectra.  The underlying tracing (small dots) is the epoch average of A; orange dashed curve is 10\,$\sigma$ smoothed photometric error per resol.  Upper panel: integrated flux, with wavelength limits indicated by hatching.  Middle panel: single Gaussian fits.  Lower panel: bimodal Gaussian fit (\ion{Si}{4}: blue and red curves for narrow and broad components, respectively) and constrained multi-line Gaussian modeling (\ion{O}{4}]\,+\,\ion{S}{4}]: green; thin blue curve represents contribution by \ion{S}{4}] $\lambda$1404).  See Appendix B for details.}
\end{figure}

\clearpage
\begin{figure}
\figurenum{5}
\vskip -10mm
\hskip -8.25mm
\includegraphics[bb= 155 98 452 693,scale=0.85,angle=90]{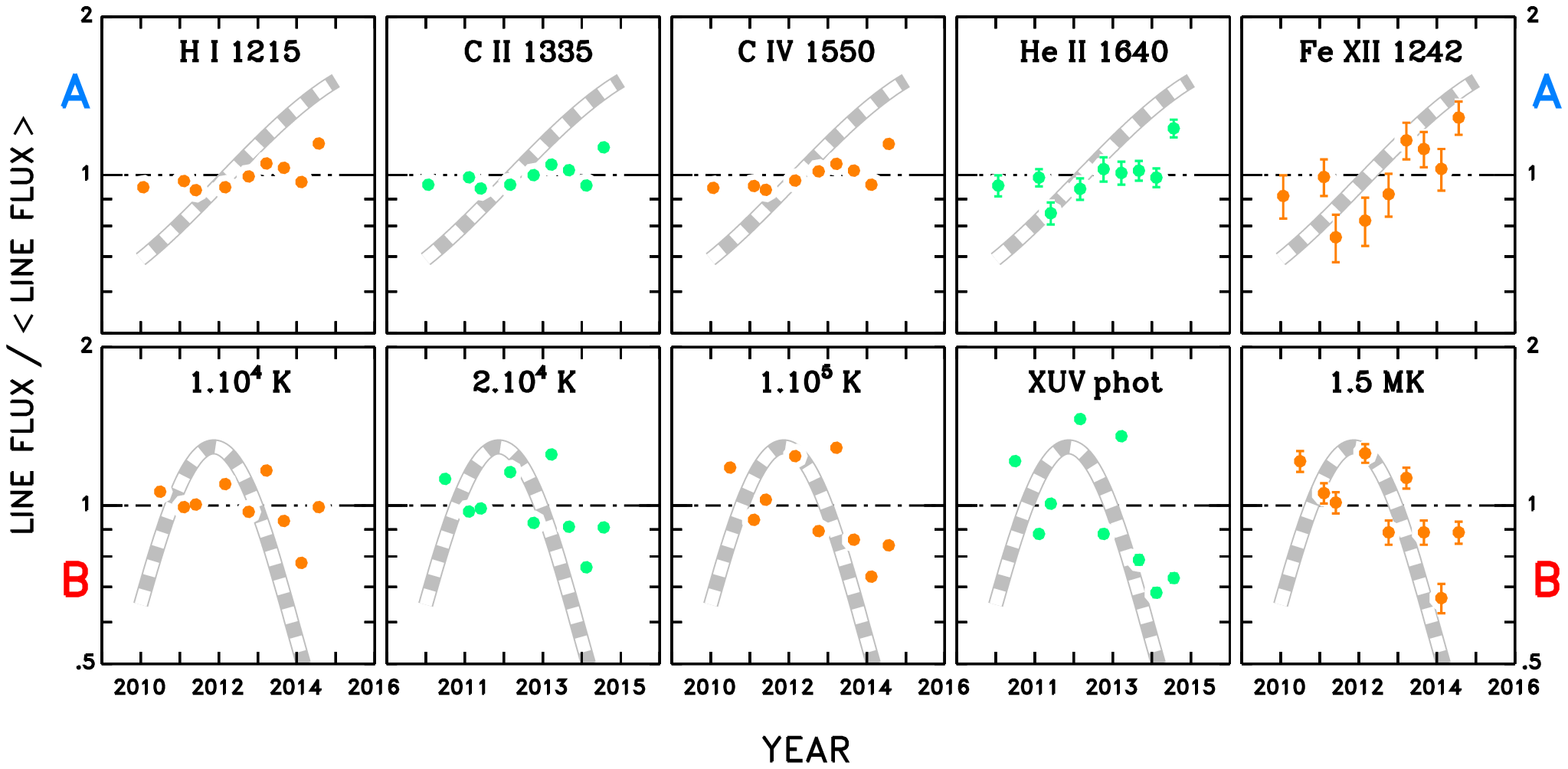} 
\vskip 0mm
\figcaption[]{Coronal cycle dependence of the integrated FUV line fluxes of various key species having different temperature sensitivities.  Light gray dashed curves are modeled X-ray cycles (Fig.~1) normalized, in same way as the FUV fluxes, to average values over the 4-year STIS period.}
\end{figure}

\clearpage
\begin{figure}
\figurenum{6a}
\vskip -10mm
\hskip 16mm
\includegraphics[bb= 70 28 509 771,scale=0.80,angle=0]{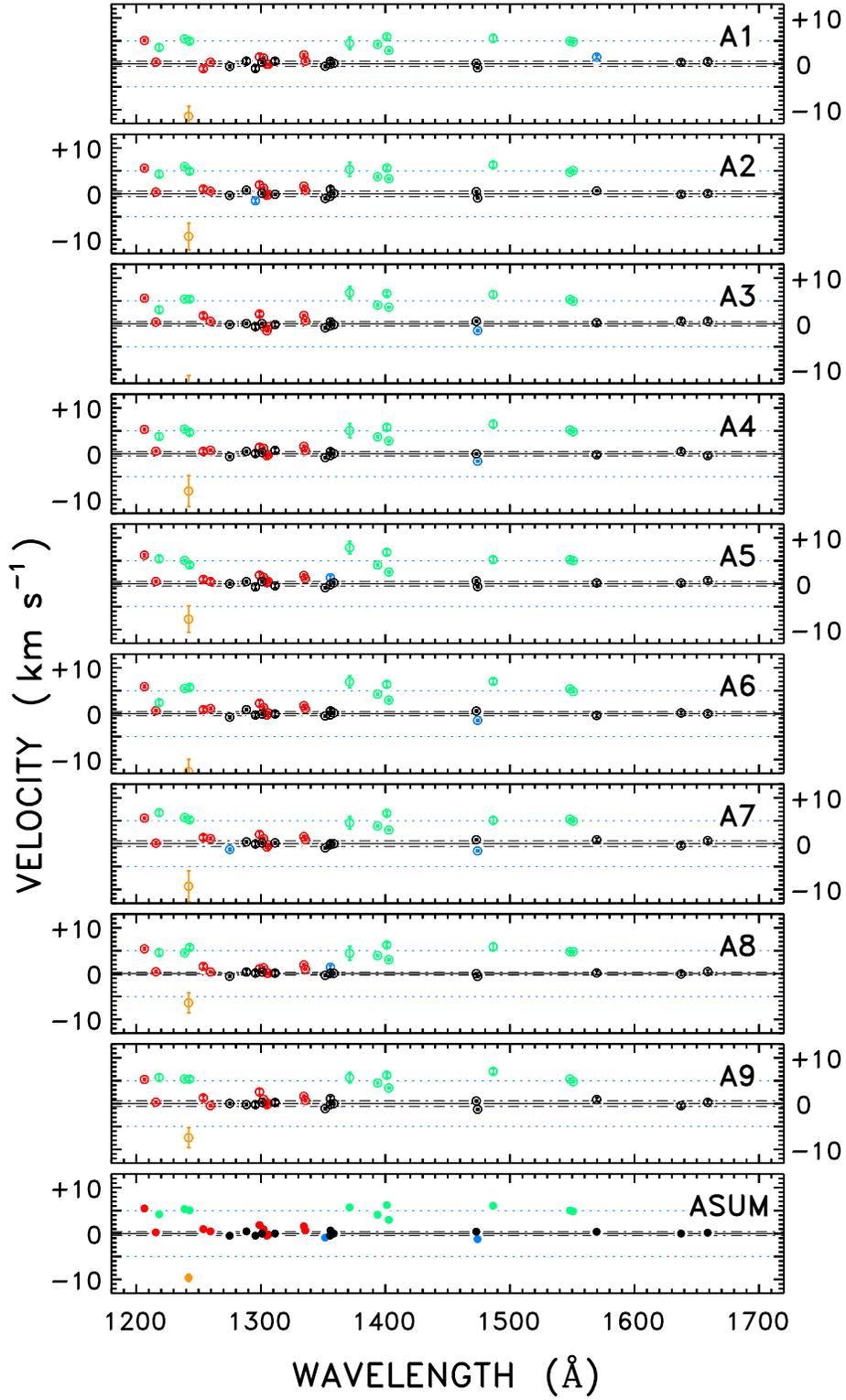} 
\vskip 0mm
\figcaption[]{Summary of single Gaussian measurements of bright, isolated FUV emission lines from the nine separate epochs, and epoch average (``SUM''), of $\alpha$~Cen A.}  
\end{figure}

\clearpage
\begin{figure}
\figurenum{6b}
\vskip -10mm
\hskip 16mm
\includegraphics[bb= 70 28 509 771,scale=0.80,angle=0]{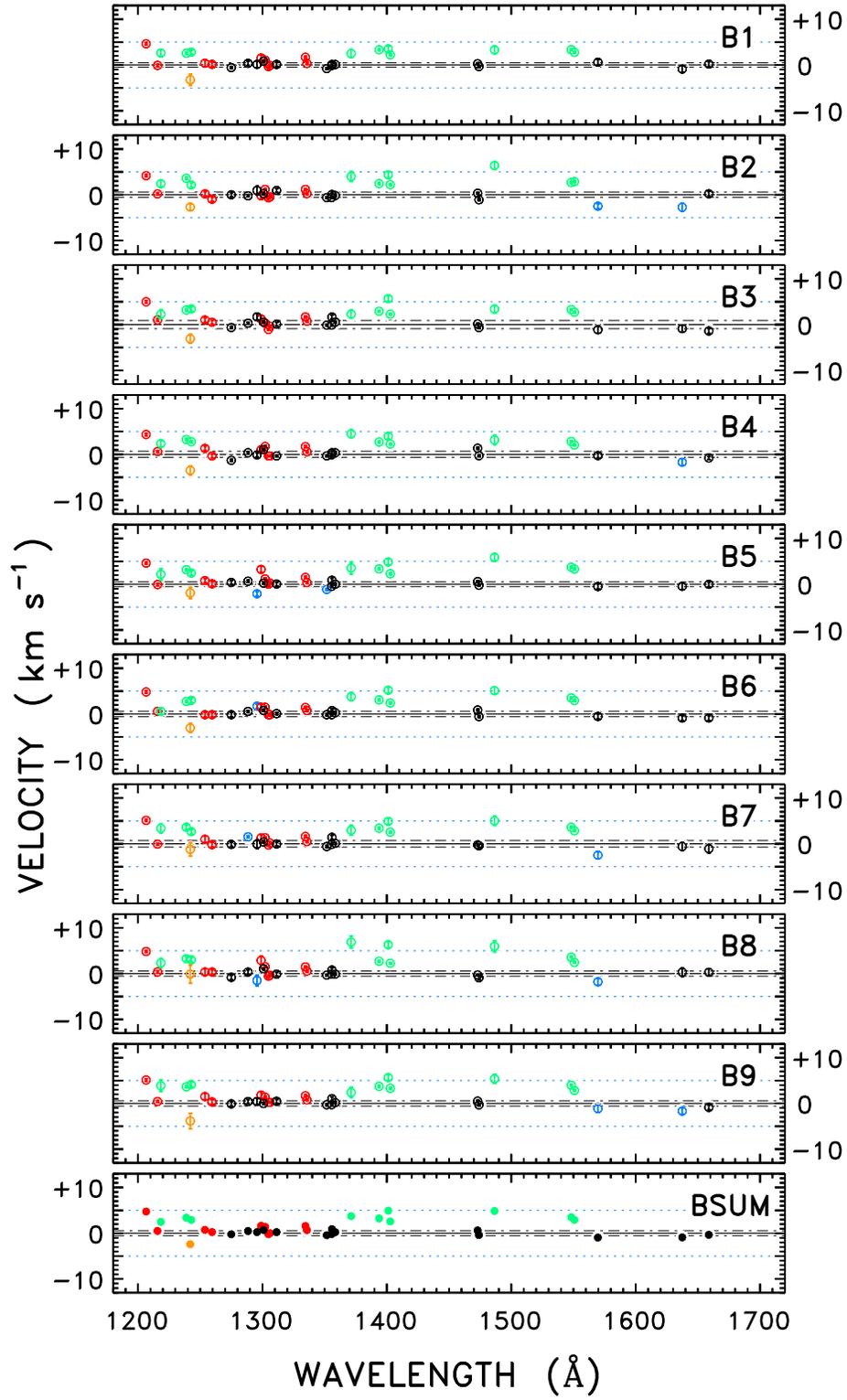} 
\vskip 0mm
\figcaption[]{Same as Fig.~6a, for $\alpha$~Cen B.}
\end{figure}

\clearpage
\begin{figure}
\figurenum{7}
\vskip -20mm
\hskip -18.5mm
\includegraphics[bb= 151 113 458 679,scale=0.95,angle=90]{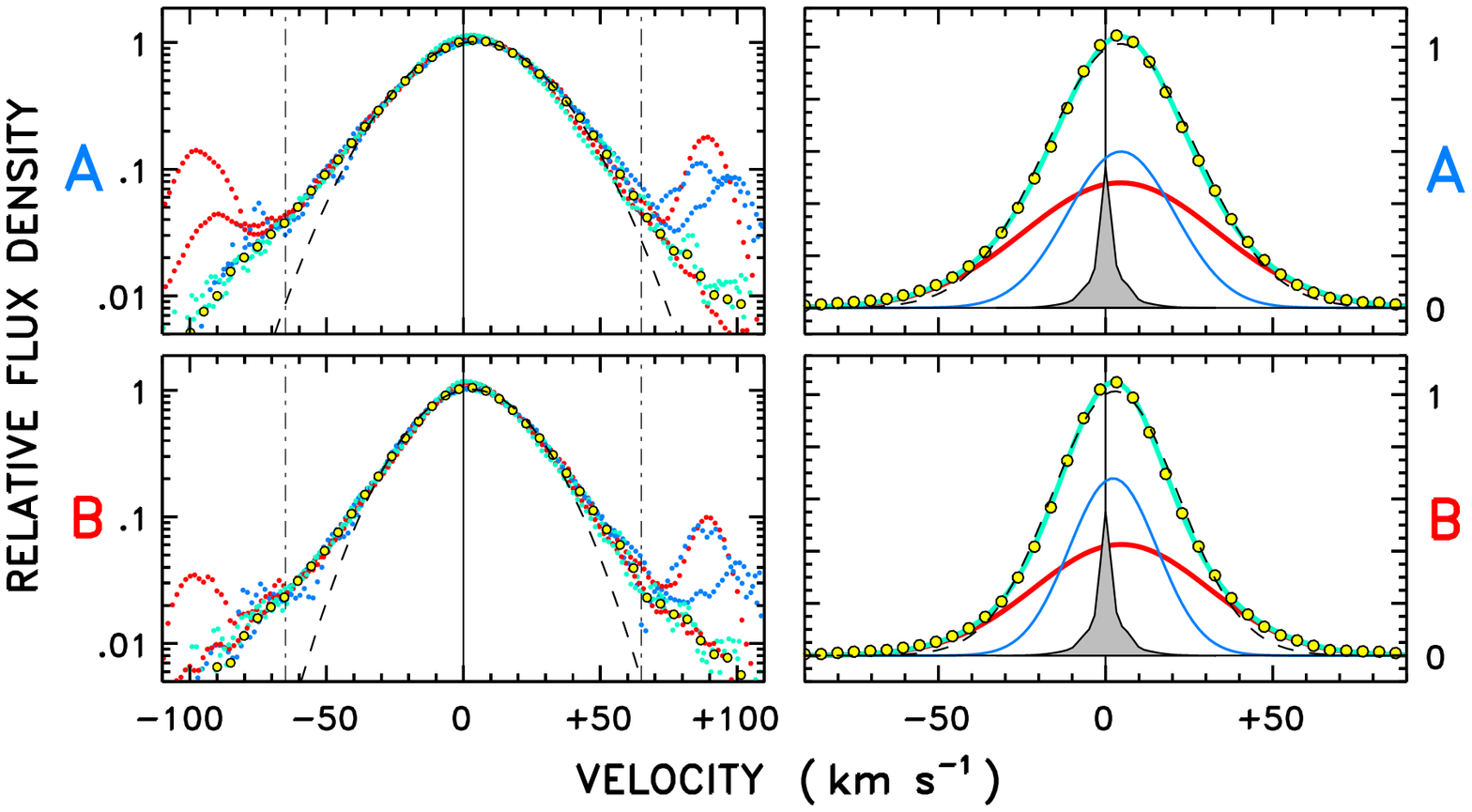} 
\vskip 0mm
\figcaption[]{Alternative, collective approach to estimating shape parameters of the TZ hot lines.  {\em Left:}\/ superposition of three hot-line doublets (smaller dots; six features total), together with filtered profile (larger dots).  Note multi-decade logarithmic flux density scale.  {\em Right:}\/ filtered profiles (large dots) and bimodal Gaussian fits, now on a linear scale: narrow and broad components are blue and red, respectively; sum is green.  On both sides of diagram, thin dashed curves are single Gaussian fits.  Gray-shaded profiles represent the E140M line-spread function for the $0.2^{\prime\prime}{\times}0.2^{\prime\prime}$ aperture.}
\end{figure}

\clearpage
\begin{figure}
\figurenum{8}
\vskip -20mm
\hskip -15mm
\includegraphics[bb= 155 98 452 693,scale=0.90,angle=90]{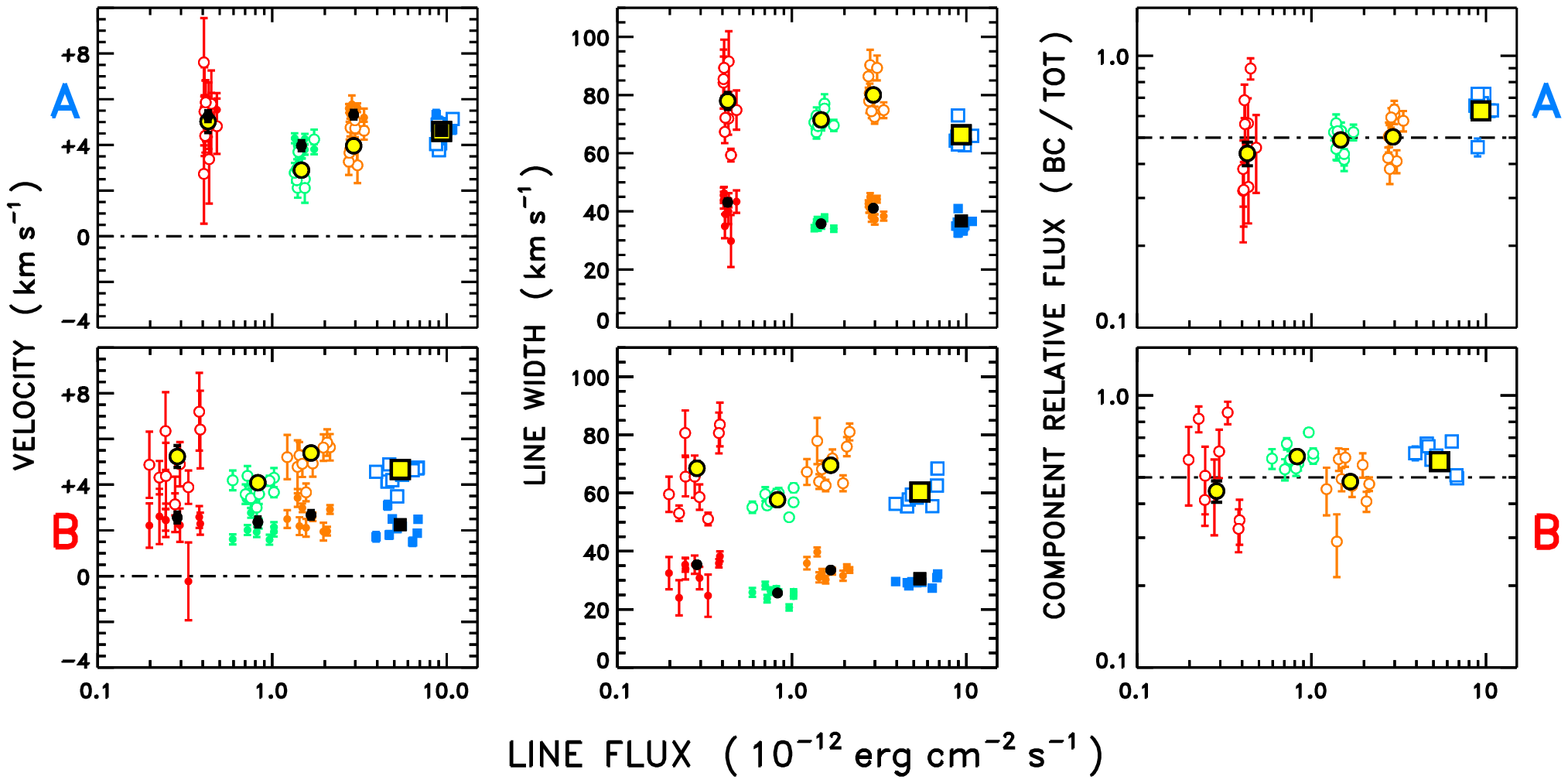} 
\vskip 0mm
\figcaption[]{Super-summary of hot-line bimodal Gaussian parameters of $\alpha$~Cen A (upper panels) and B (lower), for the individual species (circles) as well as the collective filtered profiles (squares), over the nine separate E140M epochs (smaller symbols), and for the epoch-averaged spectra (larger symbols).  Filled symbols are for the narrow components, open for the broad.  Red\,=\,\ion{N}{5}, green\,=\,\ion{Si}{4}, orange\,=\,\ion{C}{4}, and blue\,=\,\ion{Si}{4}\,+\,\ion{C}{4}\,+\,\ion{N}{5} collective profiles.}
\end{figure}

\clearpage
\begin{figure}
\figurenum{9}
\vskip -10mm
\hskip -16.5mm
\includegraphics[bb= 155 98 452 693,scale=0.95,angle=90]{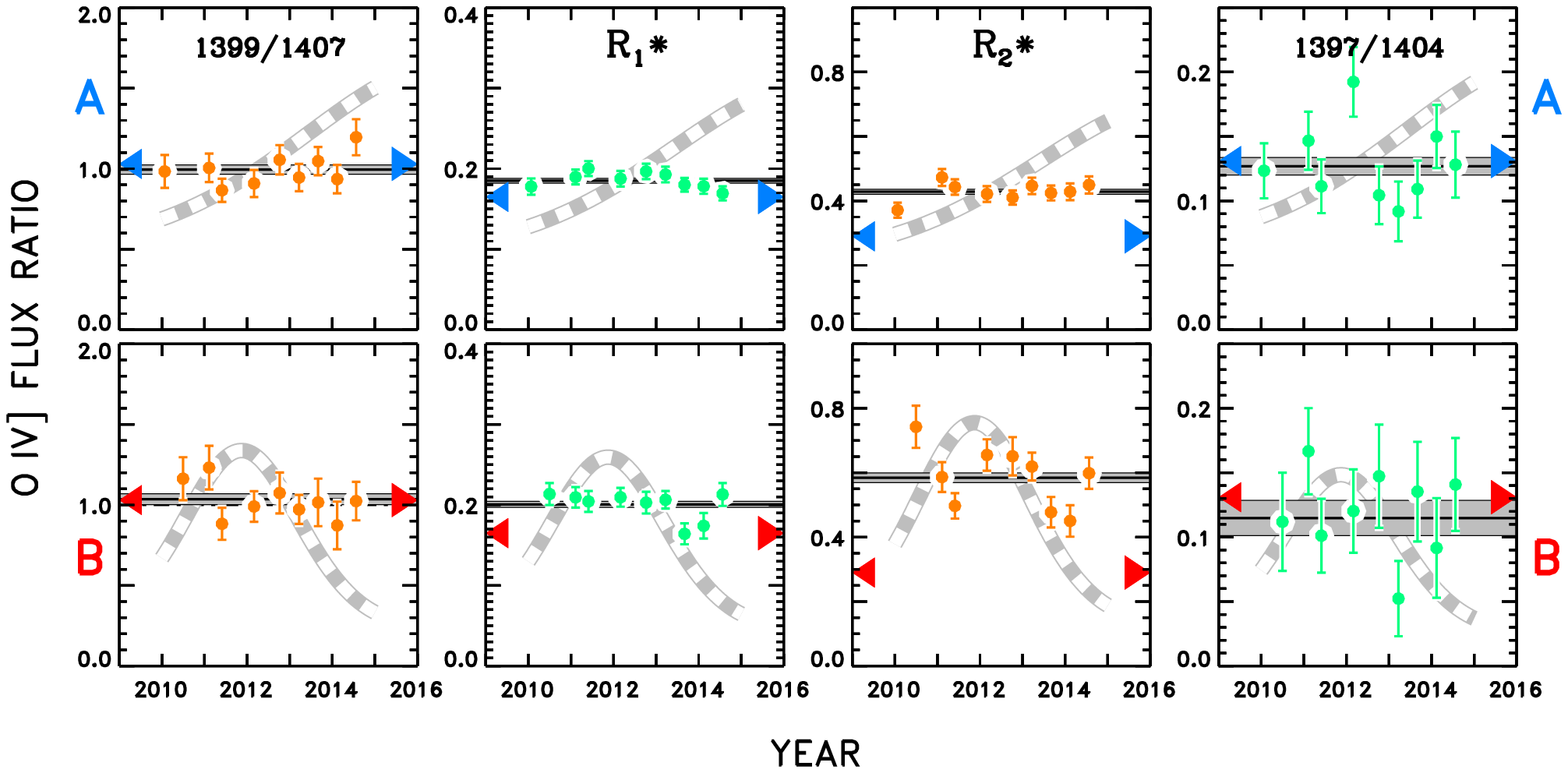} 
\vskip 0mm
\figcaption[]{Coronal cycle dependence of key \ion{O}{4}] semi-forbidden flux ratios (see text).  Arrows mark low density limits for the various ratios.  Horizontal shaded gray bars represent measurements and 1\,$\sigma$ errors from the epoch-averaged spectra.  $R_{1}{\ast}$ and $R_{2}{\ast}$ are sensitive to density over the range $n_{\rm e}= 10^{10}$--$10^{12}$ cm$^{-3}$.}
\end{figure}

\clearpage
\begin{deluxetable}{rcrcccc} 
%\rotate
\tabletypesize{\small}
\tablenum{1} 
\tablecaption{New {\em Chandra}\/ HRC Pointings} 
\tablecolumns{7}
\tablewidth{0pt} 
\tablehead{\colhead{ObsID} &  \colhead{UT Mid-Exposure} & \colhead{$t_{\rm exp}$} & 
\colhead{$({\rm CR})_{\rm A}$}  &  \colhead{$({\rm CR})_{\rm B}$}  &  
\colhead{$(L_{\rm X})_{\rm A}$}  &  \colhead{$(L_{\rm X})_{\rm B}$}  \\ 
\colhead{}   & \colhead{(yr)} & \colhead{(ks)} & \multicolumn{2}{c}{(counts s$^{-1}$)} & 
\multicolumn{2}{c}{($10^{27}$ erg~s$^{-1}$)} \\
\colhead{(1)} & \colhead{(2)} & \colhead{(3)} & \colhead{(4)} & \colhead{(5)} & \colhead{(6)} & \colhead{(7)} } 
\startdata 
   14232  &  2013.96   &  10.05  &    0.93${\pm}$0.06  &  2.00${\pm}$0.10     &     0.62   &  1.81   \\ 
   14233  &  2014.48  &    9.62   &   0.86${\pm}$0.07  &  1.39${\pm}$0.09     &     0.57   &  1.18   \\   
\enddata 
\tablecomments{Exposure time (col.~3) as reported in {\em Chandra}\/ Archive, including dead-time correction.  Count rates (cols.~4 and 5) were time-filtered to remove flare enhancements, if any; and were corrected for the 95\% encircled energy of the $r=1.6$\arcsec\ detect cell.  Uncertainties reflect standard deviations of binned count rates with respect to reported flare-filtered averages.  X-ray luminosities (0.2--2~keV; cols.~6 and 7) were derived from the count rates using source-dependent Energy Conversion Factors.  See Ayres (2014) for details.  $(L_{\rm X})_{\odot}\sim$~0.2--1.5 in same energy band and luminosity units.  $(L_{\rm bol})_{\rm A}= 1.50\,L_{\odot}$ and $(L_{\rm bol})_{\rm B}= 0.49\,L_{\odot}$ (Ayres 2009).} 
\end{deluxetable} 

\clearpage
\begin{deluxetable}{cccccc} 
%%\rotate
\tablenum{2} 
\tablecaption{{\em HST/}\,STIS Pointings 2010-2014} 
\tablecolumns{6}
\tablewidth{0pt} 
\tablehead{\multicolumn{2}{c}{Dataset} & \multicolumn{2}{c}{UT Mid-Exposure (yr)} & \multicolumn{2}{c}{Exposure Time (s)}\\[5pt]
\colhead{A} & \colhead{B} & \colhead{A} & \colhead{B} & \colhead{A} & \colhead{B}}
\startdata
ob8w01010 &   \nodata   &   2010.065   &    \nodata    &     3800    &    \nodata \\
\nodata   &  ob8w02010  &    \nodata   &    2010.497   &   \nodata   &      3600  \\
oblh01020 &  oblh01050  & \multicolumn{2}{c}{2011.108} & \multicolumn{2}{c}{4950} \\
oblh02020 &  oblh02050  &   2011.411   &    2011.412   & \multicolumn{2}{c}{4950} \\
obua01010 &  obua01020  & \multicolumn{2}{c}{2012.165} & \multicolumn{2}{c}{4275} \\
obua02010 &  obua02020  & \multicolumn{2}{c}{2012.766} & \multicolumn{2}{c}{4275} \\
oc1i10010 &  oc1i10020  & \multicolumn{2}{c}{2013.222} & \multicolumn{2}{c}{4275} \\
oc1i11010 &  oc1i11020  & \multicolumn{2}{c}{2013.669} & \multicolumn{2}{c}{4275} \\
oc7w10010 &  oc7w10020  & \multicolumn{2}{c}{2014.118} & \multicolumn{2}{c}{4275} \\
oc7w11010 &  oc7w11020  & \multicolumn{2}{c}{2014.562} & \multicolumn{2}{c}{4275} \\
\enddata 
\tablecomments{All exposures were STIS medium-resolution FUV echelle setting E140M-1425, through the 0.2$^{\prime\prime}{\times}0.2^{\prime\prime}$ aperture; split into two equal sub-exposures for $t_{\rm exp}< 4000$~s and three for $t_{\rm exp}> 4000$~s.} 
\end{deluxetable} 

\clearpage
\begin{deluxetable}{rlccl} 
\tablenum{3} 
\tablecaption{FUV Reference Wavelengths} 
\tablecolumns{4}
\tablewidth{0pt} 
\tablehead{
 \colhead{Species} & \colhead{Wavelength} & \colhead{$\lambda - \lambda_{\rm NIST}$} & \colhead{$\log{T_{\rm max}}$} & \colhead{Notes}\\[5pt]
 \colhead{}  &  \multicolumn{2}{c}{(\AA)} & \colhead{(K)} &  \colhead{} \\[5pt]
 \colhead{(1)} & \colhead{(2)} & \colhead{(3)} & \colhead{(4)} & \colhead{(5)}  
}
\startdata
%%%%%%%%%%%%%%%%%%%%%%%%%%%%%%%%%%%%%%%%%%%%%%%%%%%%%%%%%%%%%%%%%%%%%%%%%%%%%%%%%%%%%%%%%%%%%%%%%%%%%%%%%%%
        \ion{C}{3} & 1174.933  &   \nodata   &          4.8  & \\
%%%%%----------------------------------------------------------------------------------------
        \ion{C}{3} & 1175.263  &  \nodata   &               4.8  &  \\
%%%%%----------------------------------------------------------------------------------------
        \ion{C}{3} & 1175.590  &   \nodata   &              4.8  &  \\
%%%%%----------------------------------------------------------------------------------------
        \ion{C}{3} & 1175.711  &   \nodata   &              4.8  &  \\
%%%%%----------------------------------------------------------------------------------------
        \ion{C}{3} & 1175.987  &  \nodata   &               4.8  &  \\
%%%%%----------------------------------------------------------------------------------------
        \ion{C}{3} & 1176.370  &  \nodata   &               4.8  &  \\
%%%%%----------------------------------------------------------------------------------------
        \ion{Si}{3} & 1206.500  &   \nodata   &             4.7  &   \\
%%%%%----------------------------------------------------------------------------------------
        \ion{H}{1} & 1215.670  &   \nodata   &              4.3  &   \\
\enddata
\end{deluxetable} 

%%        \ion{H}{1} &     1215  & GAU1\tablenotemark{a}  & $+0.2{\pm}  0.2~(  0.2)$ & 
%%$  150{\pm}   0~(   1)$ & \nodata & $   0.229{\pm} 0.012$
%% & \nodata \\

\clearpage
\begin{deluxetable}{rlcccccc} 
\rotate
\tablenum{4a} 
\tablecaption{{\em HST/}\,STIS FUV Measurements: $\alpha$~Cen~A} 
\tablecolumns{8}
\tablewidth{0pt} 
\tablehead{
\multicolumn{2}{c}{Transition} & \colhead{Flag} & \colhead{$\upsilon_{\rm L}$} & \colhead{FWHM$_{\rm L}$} & 
\colhead{$f_{\rm L}$} & \colhead{C$_{0}$} & \colhead{C$_{1}$}\\[5pt]
 \colhead{}  & \colhead{(\AA)} & \colhead{} & \multicolumn{2}{c}{(km s$^{-1}$)} &  
\colhead{(10$^{-12}$ erg cm$^{-2}$ s$^{-1}$)} &
\multicolumn{2}{c}{(10$^{-12}$ erg cm$^{-2}$ s$^{-1}$ \AA$^{-1}$)}\\[5pt]
\colhead{(1)} & \colhead{(2)} & \colhead{(3)} & \colhead{(4)} & \colhead{(5)} & \colhead{(6)} & \colhead{(7)} & \colhead{(8)} 
}
\startdata
%%%%%%%%%%%%%%%%%%%%%%%%%%%%%%%%%%%%%%%%%%%%%%%%%%%%%%%%%%%%%%%%%%%%%%%%%%%%%%%%%%%%%%%%%%%%%%%%%%%%%%%%%%%
        \ion{C}{3} &     1174  & GAU6  & $+4.9{\pm}  0.3~(  0.8)$ & 
$   50{\pm}   0~(   1)$ & $   0.164{\pm} 0.002~( 0.010)$ & $   0.035{\pm} 0.004$
 & $   0.000{\pm} 0.000$  \\
%%%%%----------------------------------------------------------------------------------------
        \ion{C}{3} &   1175.3  & GAU6  & \nodata & \nodata & 
$   0.123{\pm} 0.002~( 0.004)$ & \nodata & \nodata  \\
%%%%%----------------------------------------------------------------------------------------
        \ion{C}{3} &   1175.6  & GAU6  & \nodata & \nodata & 
$   0.104{\pm} 0.004~( 0.018)$ & \nodata & \nodata  \\
%%%%%----------------------------------------------------------------------------------------
        \ion{C}{3} &   1175.7  & GAU6  & \nodata & \nodata & 
$   0.386{\pm} 0.004~( 0.018)$ & \nodata & \nodata  \\
%%%%%----------------------------------------------------------------------------------------
        \ion{C}{3} &   1176.0  & GAU6  & \nodata & \nodata & 
$   0.121{\pm} 0.002~( 0.008)$ & \nodata & \nodata  \\
%%%%%----------------------------------------------------------------------------------------
        \ion{C}{3} &   1176.4  & GAU6  & \nodata & \nodata & 
$   0.152{\pm} 0.002~( 0.010)$ & \nodata & \nodata  \\
%%%%%----------------------------------------------------------------------------------------
        \ion{C}{3} &     1174  & INT   & \nodata & $  547$ & 
$   1.060{\pm} 0.005~( 0.056)$ & $   0.035{\pm} 0.004$ & \nodata \\
%%%%%----------------------------------------------------------------------------------------
       \ion{Si}{3} &     1206  & GAU1  & $+5.5{\pm}  0.2~(  0.2)$ & 
$   56{\pm}   0~(   0)$ & $   1.810{\pm} 0.004~( 0.160)$ & $   0.042{\pm} 0.004$
 & \nodata \\
\enddata
\end{deluxetable} 

\clearpage
\begin{deluxetable}{rlcccccc} 
\rotate
\tablenum{4b} 
\tablecaption{{\em HST/}\,STIS FUV Measurements: $\alpha$~Cen~B} 
\tablecolumns{8}
\tablewidth{0pt} 
\tablehead{
\multicolumn{2}{c}{Transition} & \colhead{Flag} & \colhead{$\upsilon_{\rm L}$} & \colhead{FWHM$_{\rm L}$} & 
\colhead{$f_{\rm L}$} & \colhead{C$_{0}$} & \colhead{C$_{1}$}\\[5pt]
 \colhead{}  & \colhead{(\AA)} & \colhead{} & \multicolumn{2}{c}{(km s$^{-1}$)} &  
\colhead{(10$^{-12}$ erg cm$^{-2}$ s$^{-1}$)} &
\multicolumn{2}{c}{(10$^{-12}$ erg cm$^{-2}$ s$^{-1}$ \AA$^{-1}$)}\\[5pt]
\colhead{(1)} & \colhead{(2)} & \colhead{(3)} & \colhead{(4)} & \colhead{(5)} & \colhead{(6)} & \colhead{(7)} & \colhead{(8)} 
} 
\startdata
%%%%%%%%%%%%%%%%%%%%%%%%%%%%%%%%%%%%%%%%%%%%%%%%%%%%%%%%%%%%%%%%%%%%%%%%%%%%%%%%%%%%%%%%%%%%%%%%%%%%%%%%%%%
        \ion{C}{3} &     1174  & GAU6  & $+3.8{\pm}  0.3~(  0.8)$ & 
$   43{\pm}   0~(   2)$ & $  0.0835{\pm}0.0013~(0.0180)$ & $   0.014{\pm} 0.003$
 & $   0.000{\pm} 0.000$  \\
%%%%%----------------------------------------------------------------------------------------
        \ion{C}{3} &   1175.3  & GAU6  & \nodata & \nodata & 
$  0.0668{\pm}0.0013~(0.0150)$ & \nodata & \nodata  \\
%%%%%----------------------------------------------------------------------------------------
        \ion{C}{3} &   1175.6  & GAU6  & \nodata & \nodata & 
$  0.0481{\pm}0.0025~(0.0074)$ & \nodata & \nodata  \\
%%%%%----------------------------------------------------------------------------------------
        \ion{C}{3} &   1175.7  & GAU6  & \nodata & \nodata & 
$   0.214{\pm} 0.003~( 0.041)$ & \nodata & \nodata  \\
%%%%%----------------------------------------------------------------------------------------
        \ion{C}{3} &   1176.0  & GAU6  & \nodata & \nodata & 
$  0.0713{\pm}0.0014~(0.0160)$ & \nodata & \nodata  \\
%%%%%----------------------------------------------------------------------------------------
        \ion{C}{3} &   1176.4  & GAU6  & \nodata & \nodata & 
$  0.0859{\pm}0.0014~(0.0170)$ & \nodata & \nodata  \\
%%%%%----------------------------------------------------------------------------------------
        \ion{C}{3} &     1174  & INT   & \nodata & $  547$ & 
$   0.584{\pm} 0.004~( 0.110)$ & $   0.014{\pm} 0.003$ & \nodata \\
%%%%%----------------------------------------------------------------------------------------
       \ion{Si}{3} &     1206  & GAU1  & $+4.7{\pm}  0.2~(  0.2)$ & 
$   46{\pm}   0~(   0)$ & $   0.921{\pm} 0.003~( 0.130)$ & $   0.014{\pm} 0.009$
 & \nodata \\
\enddata 
\tablecomments{See Notes to Table 4a.} 
\end{deluxetable} 

%%% /Users/ayres/Documents/COPPER-AUG-2014/STIS-ALPCEN/draft/

%%%% !!!!!!!!!!!!!
\end{document}